\documentclass{jfm}
\usepackage{graphicx}
\usepackage{epstopdf, epsfig}

\usepackage{amsmath}
\usepackage{lineno}
\usepackage[normalem]{ulem}
\usepackage{soul}

\usepackage{floatrow}
\usepackage{subfig}
\usepackage{bm}

\usepackage{nameref,zref-xr}
\zxrsetup{toltxlabel}
\zexternaldocument*{Supp_Mat}

\usepackage{hyperref}
\hypersetup{
    unicode=false,          % non-Latin characters in Acrobat’s bookmarks
    pdftoolbar=true,        % show Acrobat’s toolbar?
    pdfmenubar=true,        % show Acrobat’s menu?
    pdffitwindow=false,     % window fit to page when opened
    pdfstartview={FitH},    % fits the width of the page to the window
    pdftitle={My title},    % title
    pdfauthor={Author},     % author
    pdfsubject={Subject},   % subject of the document
    pdfcreator={Creator},   % creator of the document
    pdfproducer={Producer}, % producer of the document
    pdfkeywords={keyword1, key2, key3}, % list of keywords
    pdfnewwindow=true,      % links in new PDF window
    colorlinks=true,       % false: boxed links; true: colored links
    linkcolor=blue,          % color of internal links (change box color with linkbordercolor)
    citecolor=blue,        % color of links to bibliography
    filecolor=blue,      % color of file links
    urlcolor=blue           % color of external links
}

\newcommand{\hatt}[1]{\hat{\tilde{#1}}}

\newcommand{\vct}[1]{\bm{#1}}
\newcommand{\weigh}[1]{|\vct #1|_\mathrm{w}}
\newcommand{\mtx}[1]{\bm{\mathsfbi{#1}}}

\usepackage{xcolor}
\newcommand{\inner}[2]{\big<#1\big|#2\big>}
\definecolor{burntorange}{rgb}{1, 0.49, 0.0}
\definecolor{rottengreen}{rgb}{0, 0.49, 0.0}
\definecolor{strongred}{rgb}{0.73, 0.07, 0.0}

\shorttitle{Degenerate high-order sensitivity and EPs in thermoacoustics}
\shortauthor{A. Orchini et  al.}

\title{%The role of exceptional points in thermoacoustic systems
Degenerate perturbation theory in thermoacoustics: high-order sensitivities and exceptional points}
\author{Alessandro Orchini\aff{1}
  \corresp{\email{a.orchini@tu-berlin.de}},
  Luca Magri\aff{2,3},
  Camilo F. Silva\aff{4}
  Georg A. Mensah\aff{5},
 \and Jonas P. Moeck\aff{6}}

\affiliation{\aff{1}Institute of Fluid Dynamics and Technical Acoustics, Technical University of Berlin, Berlin, Germany
\aff{2} Engineering Department, University of Cambridge, Cambridge, UK
\aff{3}Institute for Advanced Study, Technical University of Munich, Garching, Germany (visiting)
\aff{4}Department of Mechanical Engineering, Technical University of Munich, Garching, Germany
\aff{5} Department of Mechanical and Process Engineering, ETH Zurich, Zurich, Switzerland
\aff{6}Department of Energy and Process Engineering, Norwegian University of Science and  Technology, Trondheim, Norway}

\begin{document}

\maketitle

\begin{abstract}
In this study, we connect concepts that have been recently developed in thermoacoustics, specifically, (i) high-order spectral perturbation theory, (ii) symmetry induced degenerate thermoacoustic modes, (iii) intrinsic thermoacoustic modes, and (iv) exceptional points. Their connection helps gain physical insight into the behaviour of the thermoacoustic spectrum when parameters of the system are varied. First, we extend high-order adjoint-based perturbation theory of thermoacoustic modes to the degenerate case. We provide explicit formulae for the calculation of the eigenvalue corrections to any order. These formulae are valid for self-adjoint, non-self-adjoint or even non-normal systems; therefore, they can be applied to a large range of problems, including fluid dynamics. 
Second, by analysing the expansion coefficients of the eigenvalue corrections as a function of a parameter of interest, we accurately estimate the radius of convergence of the power series. Third, we connect the existence of a finite radius of convergence to the existence of singularities in parameter space. We identify these singularities as {\it exceptional points}, which correspond to defective thermoacoustic eigenvalues, with infinite sensitivity to infinitesimal changes in the parameters. At an exceptional point, two eigenvalues and their associated eigenvectors coalesce. Close to an exceptional point, strong veering of the eigenvalue trajectories is observed.
%The behaviour of the thermoacoustic eigenvalues across exceptional points can be described in terms of  Puiseux series expansions, rather than a power series. 
As demonstrated in recent work, exceptional points naturally arise in thermoacoustic systems due to the interaction between modes of acoustic and intrinsic origin. The role of exceptional points in thermoacoustic systems sheds new light on the physics and sensitivity of thermoacoustic stability, which can be leveraged for passive control by small design modifications.

\end{abstract}

\begin{keywords}
%Authors should not enter keywords on the manuscript, as these must be chosen by the author during the online submission process and will then be added during the typesetting process (see http://journals.cambridge.org	/data/\linebreak[3]relatedlink/jfm-\linebreak[3]keywords.pdf for the full list)
%Thermoacoustics instabilities \sep Perturbation theory \sep Degenerate modes \sep Exceptional point \sep Intrinsic modes
\end{keywords}

%%%%%%%%%%%%%%%%%%%%%%%%%%%%%%%%%%%%%%%%%%%%
%%%%%%%%%%%%%%%%%%%%%%%%%%%%%%%%%%%%%%%%%%%%
\section{Introduction}
\label{sec:Intro}
Thermoacoustic instability is the result of the mutual coupling between flow dynamics, the unsteady heat release produced by a flame, and the surrounding acoustic environment \citep{Dowling2003,LieuwYang05,Morgans2005,Culick2006,Poinsot2017}. Thermoacoustic  instability is a problem of major concern for the development of gas turbines that reliably work at a wide-range of operating conditions, while producing  reduced levels of carbon dioxide and NO$_\textrm{x}$ emissions that comply with environmental regulations. During thermoacoustic instability, large amplitude pressure fluctuations develop inside the combustion chamber and affect the entire engine as undesired vibrations. These vibrations affect the normal operation of the system and reduce the lifespan of the engine. In extreme cases, thermoacoustic instability may induce flashback of the flame, causing severe damage to the system elements~\citep{LieuwYang05}. Quantitative stability prediction and analysis of thermoacoustic systems require the calculation of complex-valued eigenvalues and their associated eigenvectors. Thermoacoustic eigenvalues can be found by solving a nonlinear eigenvalue problem, which is often derived from the non-homogeneous Helmholtz equation including a feedback term that represents the flame response to acoustic perturbations~\citep[e.g.,][]{NicouBenoi07}. This calculation may be computationally demanding if systems with millions of degrees of freedom are considered. In order to calculate the drift of eigenvalues and eigenvectors due to changes in parameters at an affordable computational cost, high-order adjoint-based perturbation theory can instead be used~\citep{Mensah2020}.

%%%%%%%%%%%%%%%%%%%%%%%%%%%%%%%%%%%%%%%%%%%%
\subsection{Thermoacoustic eigenvalues: classification and origin}
In this study, we consider a finite-dimensional nonlinear eigenvalue problem of the form
\begin{equation}
    \mathsfbi L(s) \hat{\bm{p}} = \bm{0},
\end{equation}
where $s$ is the eigenvalue and $\hat{\bm{p}}$ is the associated eigenvector. Nonlinear eigenproblems appear in different applications in science and engineering beyond thermoacoustics, for example, in vibrations of structures, fluid-structure interaction, nanotechnology (quantum dots), time-delay systems and control theory, to name a few~\citep{Friedman1968,Mehrmann2004,Betcke2013,Guttel2017}. The classification of the eigenvalues according to their algebraic and geometric multiplicity, and their thermoacoustic physical origin, is essential, because it reflects certain physical properties of the system, such as symmetry and sensitivity. 
In the following, we briefly recall the relevant definitions.

An eigenvalue has algebraic multiplicity $a$ if $\partial^j/\partial s^j\textrm{det}(\mathsfbi{L})=0$ and $\partial^a/\partial s^a\textrm{det}(\mathsfbi{L})\neq0$, where $j=0,1,...,a-1$.  
The geometric multiplicity, $g$, of an eigenvalue $s$ is the dimension of the null space of $\mathsfbi{L}(s)$, i.e.\ $g\equiv\dim\operatorname{null}\mathsfbi{L}(s)$.  
The geometric multiplicity is always less than or equal to the algebraic multiplicity. An eigenvalue is  semi-simple if $a=g$, it is defective if $a>g$, and it is called simple if $a=g=1$. Eigenvalues that are not simple are degenerate. Degenerate semi-simple eigenvalues are of relevance in several applications with spatial symmetries, including thermoacoustics. For example,  rotationally symmetric annular and can-annular combustors, which are are common in thermoacoustics, feature degenerate semi-simple eigenvalues. An important class of defective eigenvalues are branch-point solutions of the characteristic function, which are known as exceptional points (EPs)~\citep{Heiss2004}. As recently shown, these spectral singularities are general features of thermoacoustic systems \citep{Mensah2018_jsv,Orchini2020}. At EPs, the eigenvalues have infinite sensitivity to infinitesimal perturbations to the system. 

From a physical point of view, thermoacoustic eigenvalues can be classified according to the feedback loop between unsteady heat release and acoustics. Unsteady heat release rate generates acoustic waves, which propagate away from the flame until they reach the system's boundaries. After reflection at the boundaries, the acoustic waves impinge on the flame, modulate it, thereby generating new fluctuations in the heat release rate.
In this work, we refer to the eigenvalues associated with this feedback loop as of acoustic origin. In recent years, another feedback loop in thermoacoustic systems was discovered: the Intrinsic ThermoAcoustic (ITA) feedback loop~\citep{HoeijKorni14, BombeEmmer15}.
In the ITA loop, the upstream travelling acoustic waves produced by the flame directly modulate the upstream velocity (without reflection from the boundaries), which, in turn, causes new fluctuations in the unsteady heat release rate, which closes the loop. The ITA loop is independent of the surrounding acoustic boundaries. It exists in both anechoic environments \citep{HoeijKorni16, SilvaEmmer15} and reflective environments \citep{Emmert2017, SilvaMerk17, MukheShrir17,Orchini2020,Buschmann2020}. We refer to the eigenvalues associated with this feedback mechanism as of ITA origin.
%%%%%%%%%%%%%%%%%%%%%%%%%%%%%%%%%%%%%%%%%%%%
\subsection{Adjoint-based methods in thermoacoustics}
Thermoacoustic systems may be exceedingly sensitive to small variations in the system's parameters~\citep{Juniper2018,Magri19}. For the accurate calculation of these sensitivities, adjoint methods proved  to be efficient mathematical and  computational tools, as reviewed by~\citet{Magri19}.
Adjoint methods for thermoacoustic eigenvalue sensitivity analysis were developed for design parameter and passive control by~\citet{Magri2013}, and subsequently applied to more complex flames in \citet{Magri2013c,Orchini2016}. \citet{Rigas2016} tested experimentally adjoint-based predictions, showing that the eigenvalue shift was predicted accurately by adjoint sensitivity analysis.  The sensitivity information provided by adjoint methods can be embedded into a gradient-update optimisation  routine to optimally place and tune acoustic dampers in annular combustors~\citep{Mensah2017a}.
%%%%%%%%%%%%%%%%%%%%%%%%%%%%%%%%%%%%%%%%%%%%

The thermoacoustic eigenvalue problem is typically nonlinear in the eigenvalue $s$. Existing methodologies for the solution of nonlinear thermoacoustic eigenproblems utilize iterative schemes~\citep{NicouBenoi07}. This may be expensive, for example, for Helmholtz solvers with tens or hundreds of thousands of degrees of freedom, which makes parametric studies computationally demanding. Adjoint methods can also be exploited to simplify the solution of nonlinear eigenvalue problems. Using them, it is possible to map nonlinear eigenproblems, which are difficult to solve, into a series of linear non-homogeneous equations, which are easier to solve, to approximate eigensolutions to any desired order. For simple eigenvalues, general formulae based on the high-order expansion of the eigenvalue problem have been derived by the thermoacoustic community~\citep{Mensah2020}. The same level of generality, however, has not been reached for degenerate thermoacoustic modes, which are often found in practice due to the rotational symmetries of annular and can-annular combustors in gas turbines. In this study, higher-order eigenvalue perturbation expansions of thermoacoustic eigenvalues are extended to the degenerate case, which has challenging mathematical complications, as explained in~\S\ref{sec:pert} and~\S\ref{sec:Exc}. 
% such as the ambiguity left in the choice of a basis (a set of eigenvectors) that spans the eigenspace associated with degenerate eigenvalues. Furthermore, a given perturbation may or may not preserve the eigenvalue degeneracy, which adds a layer of complexity to analytical approaches. 

%%%%%%%%%%%%%%%%%%%%%%%%%%%%%%%%%%%%%%%%%%%%
\subsection{Exceptional points}
In the past years, the theory of exceptional points (EPs) has been widely employed to explain physical phenomena, e.g. in quantum mechanics and optics \citep{Heiss12, Miri19}.
In thermoacoustics, recent studies have shown that, in certain areas of the complex-frequency plane, small variations in a parameter of the thermoacoustic system lead to a significant change in the eigenvalues in the complex plane~\citep{SilvaPolif19,sogaro_schmid_morgans_2019}. These studies, however, did not explain what caused the observed high sensitivities. As highlighted in the present work, high sensitivity is a manifestation of the existence of EPs in the thermoacoustic spectrum. As a matter of fact, the sensitivity to infinitesimal changes in parameters is infinite at EPs \citep{Kato1980}. Practically, the existence of EPs is observed via strong veering (due to high sensitivity) of the eigenvalue trajectories in their vicinity. 

In recent work, the existence of exceptional points in the spectrum of a one-dimensional Rijke-tube has been shown~\citep{Mensah2018_jsv}. \citet{Orchini2020} extended these findings to realistic configurations, by investigating the thermoacoustic modes associated with the acoustic and ITA loop in 3D longitudinal and annular combustors with an $n$--$\tau$ flame response model. The corresponding thermoacoustic eigenvalues of acoustic and ITA origin were studied in the complex plane for systematic variations of $n$ and $\tau$. It was shown that eigenvalues of acoustic origin can coalesce with eigenvalues of ITA origin, manifesting in EPs. Furthermore, in an annular combustor, an EP may also originate from the coalescence of two eigenvalues of acoustic origin. These eigenvalues were found to be associated with two azimuthal modes, one dominant in the plenum and the other in the combustion chamber. This has analogies with the EPs arising due to the acoustic coupling between a cavity and an acoustic damper. In this respect,~\citet{Bourquard2019} experimentally demonstrated the existence of such EPs, and showed that both Helmholtz resonators and quarter-wave tube dampers achieve optimal damping performance when tuned to operate at the EPs of the closed-loop coupled acoustic system. In this study, we shall relate EPs in the spectrum of thermoacoustic systems with (i) the high sensitivity experienced by some thermoacoustic eigenvalues, and (ii) the limits of validity of high-order perturbation methods.

%%%%%%%%%%%%%%%%%%%%%%%%%%%%%%%%%%%%%%%%%%%%
\subsection{Scope}
All the concepts introduced in the introduction -- high-order perturbation theory, intrinsic thermoacoustic modes, and exceptional points --  have been independently shown to be relevant to thermoacoustic models in recent years, and thoroughly studied. They are, however, strongly interconnected. The objectives of this article are to reveal these interconnections, develop an efficient and accurate method for the calculation of thermoacoustic eigenvalue variations, and gain physical insight on the properties and behaviour of the thermoacoustic spectrum.

The article is structured as follows.
In~\S\ref{sec:pert} a general theory for high-order adjoint-based perturbation expansion of degenerate semi-simple thermoacoustic eigenvalues and eigenvectors is presented. The role of exceptional points in relation to high-order perturbation theory is discussed in~\S\ref{sec:Exc}. It is shown how exceptional points can be identified numerically, exploiting the high-order perturbation theory presented in~\S\ref{sec:pert}. Furthermore, we will discuss how knowledge on the location of EPs determines well-defined ranges of convergence for the eigenvalues estimated by perturbation theory. In~\S\ref{sec:App} we apply the presented high-order perturbation theory to two thermoacoustic cases: a simple eigenvalue of an axial combustor and a degenerate semi-simple eigenvalue of an annular combustor. Lastly, in~\S\ref{sec:defective} we discuss how perturbation theory of degenerate thermoacoustic eigenvalues can also be used at EPs by means of Puiseux series expansions. This highlights the difference between symmetry-induced degenerate modes, which are semi-simple and have finite sensitivity with respect to parameter perturbations, and degenerate modes at exceptional points, which are defective and have an infinite sensitivity.

%%%%%%%%%%%%%%%%%%%%%%%%%%%%%%%%%%%%%%%%%%%%
%%%%%%%%%%%%%%%%%%%%%%%%%%%%%%%%%%%%%%%%%%%%
\section{High-order adjoint-based perturbation theory for degenerate thermoacoustic modes}
\label{sec:pert}
 In this section, we present a general formulation of high-order adjoint-based perturbation theory for 2-fold degenerate semi-simple eigenvalues. This is the category of eigenvalues under which symmetry-induced degeneracies fall, as, for example, the thermoacoustic eigenvalues of rotationally symmetric annular combustors. With adjoint perturbation theory it is possible to (i) understand if a given perturbation unfolds the degeneracy or not; (ii) track the evolution of the split eigenvalues in the complex plane when a parameter is changed; (iii) calculate the variation of the split eigenvectors when the parameter is changed. We shall indicate with
\begin{equation}
\mathsfbi{L}(s,\varepsilon)\hat{\bm{p}} = \bm{0}
\label{eq:nep}
\end{equation}
a nonlinear eigenvalue problem that depends on a (set of) parameters $\varepsilon$. $\mathsfbi{L}$ is a linear operator acting on an eigenvector $\hat{\bm{p}}$. The pairs $(s,\hat{\bm{p}})$ for which~\eqref{eq:nep} is satisfied represent the eigenvalues and eigenvectors of the operator. The operator $\mathsfbi{L}$ is assumed to have an analytical dependence on the eigenvalue and the parameter(s). No further assumptions are made on the properties of the operator $\mathsfbi{L}$, which, in general, can be non-self-adjoint or even non-normal.
Its corresponding adjoint operator, $\mathsfbi{L}^\dagger$, is defined via
\begin{equation}
\inner{\bm{g}}{\mathsfbi{L} \bm{f}} \equiv \inner{\mathsfbi{L}^\dagger \bm{g}}{\bm{f}},
\label{eq:defAdj}
\end{equation}
where $\inner{\cdot}{\cdot}$ is an inner product, and $\bm{f}$ and $\bm{g}$ are arbitrary complex-valued vectors in their relevant Hilbert spaces. In the following, we shall adopt the Hermitian inner product $\inner{\bm{g}}{\bm{f}} = \bm{g}^H \bm f$, where the superscript $^H$ indicates conjugate transpose. Note that, according to this definition, the direct and adjoint operators have the same eigenvalues~\citep{Lopez2007,Guttel2017} and, in a discretized finite element framework as that used in~\S\ref{sec:App}, the discrete adjoint operator is equivalent to the Hermitian transpose of the direct operator, i.e.,  $\mathsfbi{L}^\dagger=\mathsfbi{L}^H$. For the thermoacoustic problem, the eigenproblem that we are solving is~\citep{Dowling2003,Culick2006,NicouBenoi07}
\begin{equation}
\left[\nabla\bm\cdot(c^2\nabla) - s^2 -\frac{(\gamma - 1)}{\overline{\rho}} \frac{\overline{Q}}{\overline{U}}n e^{-s\tau}\hat{\bm{n}}_{\mathrm{ref}} \bm{\cdot} \nabla_{\mathrm{ref}}\right]\hat{p}=0,% \quad \mbox{and b.c.},
\label{eq:nep_ta}
\end{equation}
coupled with a set of boundary conditions. 
The finite-dimensional operator $\mathsfbi{L}$ and the eigenvector $\hat{\bm{p}}$ in~\eqref{eq:nep} are, respectively, the discretisation of the thermoacoustic operator and the acoustic pressure $\hat{p}$ in~\eqref{eq:nep_ta}. In Eq.~\eqref{eq:nep_ta}, which is valid in the zero mean Mach number limit, $c$ is the speed of sound (which may vary spatially), $\gamma$ is the ratio of specific heats, assumed to be homogeneous, and $\overline{\rho}$, $\overline{Q}$, $\overline{U}$ are the mean density, heat release rate and flow velocity, respectively. The last term on the l.h.s. represents the effect of unsteady heat release on the acoustics, modelled with a so-called $n$--$\tau$ model~\citep{Crocco1965}. According to this model, the flame response is proportional to the delayed axial acoustic velocity fluctuations at a reference location, $_{\mathrm{ref}}$, upstream of the flame. Together with non-trivial boundary conditions~\citep{NicouBenoi07}, the flame response causes the thermoacoustic operator $\mathsfbi{L}$ to have non-orthogonal eigenvectors, thus exhibiting a non-normal response. Additionally, the delayed response of the flame also causes the thermoacoustic operator to be nonlinear in the eigenvalue $s$. In the present study, we will consider both flame response coefficients, $n$ and $\tau$, as perturbation parameters. We highlight that the use of an $n$--$\tau$ model is not a limitation of the perturbative method that we will discuss. The method can be applied to any flame model that is analytic in the eigenvalue. This was demonstrated in~\citet{Mensah2018_jtp_nonsym} on the basis of an experimentally measured flame response expressed in state-space form.

For semi-simple eigenvalues, the eigenvalue and eigenvector dependence on a parameter $\varepsilon$ is expressed in terms of power series expansions of the form  
\begin{equation}
s(\varepsilon) \approx s_0 + \sum_{j=1}^N \varepsilon^j s_j, \qquad \qquad \bm{\hat p}(\varepsilon) \approx \bm{\hat p}_0 + \sum_{j=1}^N \varepsilon^j \bm{\hat p}_j,
\label{eq:psexp}
\end{equation}
where, without loss of generality, the perturbation parameter $\varepsilon$ is centred at a reference value $\varepsilon_0=0$. The coefficients $s_j$ and $\hat{\bm{p}}_j$ are the $j$th-order corrections to the eigenvalues and eigenvectors, respectively. The approximation symbols in \eqref{eq:psexp} indicate that the power series are truncated at order $N$. In thermoacoustics, arbitrarily high-order adjoint-based perturbation theory for non-degenerate eigenvalues has been presented in~\cite{Mensah2020}. We report here the key ideas and results of the method since they serve as a starting point for the discussion of the degenerate case, which is the main focus of this study.
It is convenient to define the shorthand
\begin{equation}
\mathsfbi{L}_{n,m} \equiv \frac{1}{n!m!}\left.\frac{\partial^{n+m} \mathsfbi{L}}{\partial s^n\partial \varepsilon^m}\right|_{\substack{s=s_0\\\varepsilon=0}}. 
\label{eq:shorthand}
\end{equation}
The power series approximations~\eqref{eq:psexp} are substituted into the eigenvalue problem~\eqref{eq:nep}, which is then expanded into a Taylor series. 
By collecting the terms at every order of $\varepsilon$, one obtains a series of linear, non-homogeneous equations that need to be solved in ascending order
\begin{equation}
\mathsfbi{L}_{0,0}\hat{\bm{p}}_j = -\bm{r}_j -s_j\mathsfbi{L}_{1,0}\hat{\bm{p}}_0, \qquad  \mbox{for } j=1,\ldots,N.
\label{eq:simPert}
\end{equation}
We refer to~\cite{Mensah2020} for a detailed derivation of Eq.~\eqref{eq:simPert}. The complexity of the equations is hidden in the $\bm{r}_j$ terms, which (i) contain all the possible ways of distributing $j$ derivatives between $s$, $\varepsilon$ and $\hat{\bm{p}}$, and (ii) are functions of the eigenvalue and eigenvector corrections $s_k$ and $\hat{\bm{p}}_k$ at orders $k<j$. Explicit expressions for the list of all the terms that compose $\bm{r}_j$ at any order can be analytically obtained. This helps the recursive implemention of perturbation theory~\citep{Mensah2020}. In Appendix~\ref{app:A}, we provide a general formula for $\bm{r}_j$ at any order, and its explicit expressions for $j=1,2$.

The solution strategy becomes straightforward: at any order, a solvability condition based on the Fredholm alternative is imposed, by projecting the r.h.s. of Eq.~\eqref{eq:simPert} onto the adjoint eigenvector $\hat{\bm{p}}_0^\dagger$, defined by $\mathsfbi{L}_{0,0}^H\hat{\bm{p}}_0^\dagger=\bm{0}$. This yields a general equation for the eigenvalue correction at order $j$
\begin{equation}
s_j = - \frac{\inner{\hat{\bm{p}}_0^\dagger}{\bm{r}_j}}{\inner{\hat{\bm{p}}_0^\dagger}{\mathsfbi{L}_{1,0}\hat{\bm{p}}_0}}.
\label{eq:nonDeg}
\end{equation}
Note that, at first order, for which $\bm{r}_1 = \mathsfbi{L}_{0,1}\hat{\bm{p}}_0$, one finds $s_1 = -\inner{\hat{\bm{p}}_0^\dagger}{\mathsfbi{L}_{0,1}\hat{\bm{p}}_0}/\inner{\hat{\bm{p}}_0^\dagger}{\mathsfbi{L}_{1,0}\hat{\bm{p}}_0}$, retrieving the known first-order sensitivity expression for nonlinear eigenvalue problems~\citep{Magri2016b}. Once the eigenvalue correction at order $j$ is known, it can be substituted back into the linear systems~\eqref{eq:simPert}, which can then be solved with standard methods for $\hat{\bm{p}}_j$. Although its solution is not unique -- \eqref{eq:simPert} is an underdetermined system of equations -- the ambiguity in the solution can always be addressed by choosing a normalization condition for the eigenvectors. With both eigenvalue and eigenvector corrections at order $j$, we can move to order $j+1$ and repeat the procedure, up to any desired order.

High-order expressions of thermoacoustic eigenvalue sensitivities have not been developed for the degenerate case. The state-of-the-art is a second-order analysis for the eigenvalues only~\citep{Magri2016b,Mensah2018_jtp_nonsym}. Starting from the procedure outlined above, in the following we show how perturbation theory can be generalised to handle degenerate semi-simple eigenvalues. We will show that, in order to develop a theory generalisable to arbitrarily high order, perturbation theory of degenerate semi-simple eigenvalues needs to be carried out in parallel on both members of the degenerate eigenvalue pair.

%%%%%%%%%%%%%%%%%%%%%%%%%%%%%%%%%%%%%%%%%%%%
\subsection{Baseline and adjoint degenerate solution}
As in any perturbative method, we first require a baseline solution. We shall assume that the baseline solution, obtained for $\varepsilon = 0$, is degenerate with algebraic multiplicity 2 and semi-simple, so that the geometric multiplicity is also 2. We thus have an unperturbed eigenvalue $s_0$ with an associated 2-dimensional subspace spanned by two eigenvectors, denoted $\hatt{\bm{p}}_{0,1}$ and $\hatt{\bm{p}}_{0,2}$, which are chosen to be orthonormal without loss of generality. The first subscript refers to the expansion order, and the second to distinguish between the modes in the degenerate eigenvalue. The tilde symbol highlights that the choice of these vectors is not unique. We also need to calculate the associated adjoint eigenvectors, $\hatt{\bm{p}}^\dagger_{0,1}$ and $\hatt{\bm{p}}^\dagger_{0,2}$, which satisfy $\mathsfbi{L}_{0,0}^H\hatt{\bm{p}}^\dagger_{0,\zeta}=\bm{0}$, for $\zeta = 1,2$.  As a convention, the subscripts of the following equations will contain Latin letters to indicate the perturbation order, and Greek letters to distinguish between the (two) degenerate modes.

For semi-simple eigenvalues, it can be shown that the direct and adjoint eigenvectors can always be chosen to satisfy the bi-orthonormalization condition~\citep{Guttel2017}
\begin{equation}
\inner{\hatt{\bm{p}}^\dagger_{0,\zeta}}{\mathsfbi{L}_{1,0} \hatt{\bm{p}}_{0,\eta}} = \delta_{\zeta,\eta},
\label{eq:ortoBio}
\end{equation}
with $\mathsfbi{L}_{1,0}$ defined via~\eqref{eq:shorthand}. This condition is valid also for non-normal operators, and we will adopt it to simplify the perturbative equations. 

%%%%%%%%%%%%%%%%%%%%%%%%%%%%%%%%%%%%%%%%%%%%
\subsection{Solvability conditions}
\label{sec:solvCond}
Because the operator $\mathsfbi{L}_{0,0}$ has a nullspace of dimension two (spanned by $\hatt{\bm{p}}_{0,\zeta}$), each of the perturbative equations~\eqref{eq:simPert} requires two solvability  conditions. More specifically, for the equations to admit solutions, their r.h.s. must be orthogonal to the (two-dimensional) adjoint subspace spanned by $\hatt{\bm{p}}^\dagger_{0,\zeta}$.  Depending on whether eigenvalue splitting has occurred or not, different solutions strategies need to be employed. This is discussed in the following.

%%%%%%%%%%%%%%%%%%%%%%%%%%%%%%%%%%%%%%%%%%%%
\subsubsection{Case 1: Degeneracy is not resolved}
\label{sec:degNRes}
As long as the perturbation considered does not resolve the degeneracy (e.g., for perturbations that preserve the symmetry of the problem), the perturbed eigenvalues will remain degenerate, and the ambiguity in the choice of a basis in the nullspace of the perturbed operator will persist. Thus, at an arbitrary order $j$, we have that the two eigenvalue corrections at orders $0\leq k<j$ are identical, $s_{k,1} = s_{k,2} = s_k$, and the degenerate subspace correction is given by the linear combination
\begin{equation}
\hat{\bm{p}}_k = \alpha_1 \hatt{\bm{p}}_{k,1} + \alpha_2 \hatt{\bm{p}}_{k,2},
\label{eq:subSp}
\end{equation}
where the $\alpha_\zeta$ coefficients are, without loss of generality, chosen to be identical at every order $k$. We are therefore still tracking one degenerate eigenvalue, governed by equation~\eqref{eq:simPert}. By imposing the two solvability conditions at order $j$, we obtain
\begin{subequations}
\begin{align}
\inner{\hatt{\bm{p}}_{0,1}^\dagger}{\bm{r}_j} + \inner{\hatt{\bm{p}}_{0,1}^\dagger}{s_j\mathsfbi{L}_{1,0}\hat{\bm{p}}_0} &= 0,\\
\inner{\hatt{\bm{p}}_{0,2}^\dagger}{\bm{r}_j} + \inner{\hatt{\bm{p}}_{0,2}^\dagger}{s_j\mathsfbi{L}_{1,0}\hat{\bm{p}}_0} &= 0.
\end{align}
\label{eq:2Conds}%
\end{subequations}
Each of the terms contained in $\bm{r}_j$ is proportional to $\hat{\bm{p}}_k$ for some $k<j$, which can be expressed as~\eqref{eq:subSp}. By indicating with $\tilde{\bm{r}}_{j,\zeta}$ the terms of $\bm{r}_j$ proportional to $\hatt{\bm{p}}_{k,\zeta}$, the solvability conditions~\eqref{eq:2Conds} can be written in matrix form as
\begin{equation}
\bm{X}_j \bm{\alpha} = s_j \bm\alpha,
\label{eq:auxPs}
\end{equation}
where $\bm{X}_{j_{\eta,\zeta}} \equiv - \inner{\hatt{\bm{p}}_{0,\eta}^\dagger}{\tilde{\bm{r}}_{j,\zeta}}$, and $\bm\alpha \equiv [\alpha_1,\alpha_2]^T$. Equation~\eqref{eq:auxPs} is a 2$\times$2 linear eigenvalue problem, which we refer to as the auxiliary eigenvalue problem. We need to distinguish two solution cases:
\begin{enumerate}
\item if the two eigenvalues of~\eqref{eq:auxPs} are identical, the problem remains degenerate at this order. We therefore cannot uniquely determine a basis for the eigenvector corrections, but it is convenient to choose them as the solutions of the linear equations
\begin{equation}
\mathsfbi{L}_{0,0}\hatt{\bm{p}}_{j,\zeta} = -\tilde{\bm{r}}_{j,\zeta} - s_j\mathsfbi{L}_{1,0} \hatt{\bm{p}}_{0,\zeta}, \qquad \mbox{for } \zeta=1,2,
\label{eq:degdeg}
\end{equation}
so that equation~\eqref{eq:subSp} holds also at order $j$, and, at the next order, the same procedure outlined in this subsection can be applied;
\item  if the two eigenvalues of~\eqref{eq:auxPs} are different, the degeneracy unfolds at this order. Together with the eigenvalue corrections $s_{j,\zeta}$, which have different values, we obtain the eigenvectors $\bm\alpha_\zeta$ that uniquely determine the directions along which the degenerate subspace of the problem unfolds at lower orders as
\begin{equation}
\hat{\bm{p}}_{k,\zeta} = [\hatt{\bm{p}}_{k,1}, \hatt{\bm{p}}_{k,2}] \bm\cdot \bm\alpha_\zeta, \qquad \mbox{for } k = 0,\ldots,j-1.
\end{equation}
This is the appropriate basis with which to investigate the problem at higher orders because, from~\eqref{eq:psexp}, it ensures that $\hat{\bm{p}}_\zeta(\varepsilon)$ smoothly approaches $\hat{\bm{p}}_{0,\zeta}$ when $\varepsilon\rightarrow 0$. To each eigenvalue at order $j$ corresponds an eigenvector correction $\hat{\bm{p}}_{j,\zeta}$ defined by
\begin{equation}
\mathsfbi{L}_{0,0}\hat{\bm{p}}_{j,\zeta} = -\bm{r}_{j,\zeta} - s_{j,\zeta}\mathsfbi{L}_{1,0} \hat{\bm{p}}_{0,\zeta}, \qquad \mbox{for } \zeta=1,2.
\label{eq:degsim}
\end{equation}
Note the differences between~\eqref{eq:degdeg} and~\eqref{eq:degsim}: in the latter the tilde symbols have been dropped because the basis is uniquely determined, and an additional index has been appended to the eigenvalue correction at order $j$, as the two eigenvalues now follow different trajectories.
\end{enumerate}

%There is one last important remark that applies to both cases just discussed.
Importantly, the system of equations for the eigenvector corrections,~\eqref{eq:degdeg} or~\eqref{eq:degsim}, admits solutions but is underdetermined since the matrix~$\mathsfbi{L}_{0,0}$ has a non-zero nullspace. Therefore, it admits an infinite number of solutions, which can be expressed as
\begin{equation}
\hat{\bm{p}}_{j,\zeta} = \hat{\bm{p}}^\bot_{j,\zeta} + c_{j,\zeta,1} \hat{\bm{p}}_{0,1} + c_{j,\zeta,2} \hat{\bm{p}}_{0,2}, 
\label{eq:Fullp}
\end{equation}
where $\hat{\bm{p}}^\bot_{j,\zeta}$ is orthogonal to the unperturbed degenerate subspace, and $c_{j,\zeta,\eta}$ are undetermined coefficients [there are two coefficients ($\eta$) for each order ($j$) for each eigenvalue ($\zeta$)]. As for the non-degenerate case, one coefficient associated with each eigenvector can be determined by imposing a normalization condition on the eigenvectors. The other coefficients, however, are uniquely determined by solvability conditions at higher orders if the eigenvalues split; this will be discussed in \S\ref{sec:degRes}.

%%%%%%%%%%%%%%%%%%%%%%%%%%%%%%%%%%%%%%%%%%%%
\subsubsection{Case 2: Degeneracy is resolved}
\label{sec:degRes}
If at a certain order the perturbation resolves the degeneracy, the eigenvalues split, and we can identify the unique eigendirections along which this splitting occurs. Let us assume that the degeneracy is resolved at order $d$. At orders $n>d$, we are therefore tracking two branches (solutions), whose equations are governed by~\eqref{eq:degsim}, and have as unknowns two eigenvalue and two eigenvector corrections. The four solvability conditions (two for each branch) in this case read
\begin{equation}
\inner{\hat{\bm{p}}_{0,\eta}^\dagger}{\bm{r}_{j,\zeta}} + \inner{\hat{\bm{p}}_{0,\eta}^\dagger}{s_{j,\zeta}\mathsfbi{L}_{1,0}\hat{\bm{p}}_{0,\zeta}} = 0, \qquad \mbox{for } \zeta=1,2 \mbox{ and } \eta = 1,2.
\label{eq:2Conds2}
\end{equation}
By exploiting the bi-orthonormality condition~\eqref{eq:ortoBio}, these reduce to
\begin{subequations}
\begin{align}
&s_{j,\zeta} = -\inner{\hat{\bm{p}}_{0,\zeta}^\dagger}{\bm{r}_{j,\zeta}}  &\mbox{if } \eta=\zeta, 
\label{eq:Fred1}\\
&\inner{\hat{\bm{p}}_{0,\eta}^\dagger}{\bm{r}_{j,\zeta}} = 0  &\mbox{if } \eta \neq \zeta.
\label{eq:Fred2}
\end{align}
\label{eq:2Conds3}%
\end{subequations}
The solvability condition~\eqref{eq:Fred1} defines the eigenvalue corrections on each branch $\zeta$ at order $j$, and is identical to the non-degenerate equation~\eqref{eq:nonDeg} when the bi-orthonormalization condition~\eqref{eq:ortoBio} is considered. The second condition, \eqref{eq:Fred2} instead, is new, and belongs to the degenerate case only. It has not been considered by the thermoacoustic community so far, which is why the current state-of-the-art on perturbation theory~\citep{Magri2016b} is limited to second order. If not considered, the solvability conditions are not satisfied, which would then lead to incorrect results in the evaluation of the eigenvector corrections and higher-order coefficients. This fact was first mentioned by~\cite{Mensah2018_jtp_nonsym} and is formally demonstrated in a complete form in the current study.

The degrees of freedom that can be leveraged to satisfy the conditions~\eqref{eq:Fred2} are the coefficients $c_{j-d,\zeta,\eta}$. In fact, $\bm{r}_{j,\zeta}$ is a function of all the eigenvector corrections $\hat{\bm{p}}_{k,\zeta}$ that have been determined at orders $k<j$, and due to~\eqref{eq:Fullp}, it is a function of all the coefficients $c_{k,\zeta,\eta}$. Analogous to the derivation outlined by~\cite{Mensah2020}, it can be shown that all the coefficients with $k>j-d$ have no influence on the order $j$ conditions $\eqref{eq:2Conds3}$, and that the order $j-d$ coefficients that guarantee solvability are given by
\begin{equation}
c_{j-d,\zeta,\eta} = \frac{\inner{\hat{\bm{p}}^\dagger_{0,\eta}}{\bm{r}_{j,\zeta}^\bot}}{s_{d,\eta}-s_{d,\zeta}} \qquad \mbox{for } \eta\neq \zeta,
\label{eq:coeffCorr}
\end{equation}
where the terms in $\bm{r}_{j,\zeta}^\bot$ include all the information available at order $j$ on the eigenvectors $\hat{\bm{p}}_{k,\zeta}$ -- specifically, the orthogonal components  $\hat{\bm{p}}^\bot_{k,\zeta}$ and all the coefficients $c_{k,\zeta,\eta}$ for $k<j-d$. A derivation of this equation in the case $d=1$, which is the most common scenario, is outlined in Appendix~\ref{app:B}. The general case is treated in \S\ref{sec:SuppDer} of the Supplementary Material. Once both the eigenvalue corrections $s_{j,\zeta}$ and the coefficients $c_{j,\zeta,\eta}$ have been evaluated, Eq.~\eqref{eq:degsim} is guaranteed to be solvable, the eigenvector corrections $\hat{\bm{p}}_{j,\zeta}$ can be calculated, and one can finally proceed to the next order. 

Equation~\eqref{eq:coeffCorr} is a theoretical contribution of this study, and is important for several reasons. It is inversely proportional to the eigenvalue split gap that occurred at order $d$; the numerator is formally equivalent to the eigenvalue correction equation, but with the adjoint eigenvector chosen to be that of the ``other'' branch ($\eta\neq \zeta$); although it is obtained at order $j$, it contains no unknown terms at this order, and instead it defines coefficients at order $j-d$. This is consistent with the fact that the numerator is of order $\mathcal{O}(\varepsilon^j)$, whereas the  denominator is of order $\mathcal{O}(\varepsilon^d)$. As a consequence, in order to obtain perturbations accurate to order $N$, an expansion at order $N+d$ is needed.
By repeatedly applying the equations contained in~\S\ref{sec:degNRes} and~\S\ref{sec:degRes}, one can calculate the eigenvalue and eigenvector coefficients of power series expansions of degenerate, semi-simple eigenvalues to arbitrarily high orders.

To conclude this theoretical section, we highlight that, in most cases of practical relevance, perturbations unfold degenerate states at first order ($d=1$). This is known as complete regular splitting~\citep{Lancaster2003}. The solution of the first-order equations ($j=1$) then follows what is described in~\S\ref{sec:degNRes} and determines the eigenvalue splitting and the correct basis. At second order ($j=2$), the solution follows what is described in~\S\ref{sec:degRes}, from which one can see that the expression for the eigenvalues is still exact (because no coefficients $c$ are evaluated at first order). However, also the coefficients $c_{1,\zeta,\eta}$ need to be determined for solvability at second order; if these are ignored, all the higher-order coefficients for both eigenvalues and eigenvectors will be incorrect. Perturbations that unfold the degeneracy at first order were discussed by~\cite{Magri2016b}, where only variations in the eigenvalues and not in the eigenvectors were considered; this explains why the perturbation theory that was outlined in that study was applicable up to second order only.

%%%%%%%%%%%%%%%%%%%%%%%%%%%%%%%%%%%%%%%%%%%%
%%%%%%%%%%%%%%%%%%%%%%%%%%%%%%%%%%%%%%%%%%%%
\section{Radius of convergence and exceptional points}
\label{sec:Exc}
The theory introduced in~\S\ref{sec:pert} yields approximations for the parametric dependence of simple and semi-simple eigenvalues and their associated eigenvectors. It provides explicit expressions for the coefficients of power series expansion up to arbitrary order. For a simple eigenvalue $s$, the function $s=s(\varepsilon)$ can always be locally expanded into a power series up to arbitrary order~\citep{Kato1980}. For degenerate semi-simple eigenvalues, power series expansions at high orders can also generally be obtained, provided that the eigenvalue splitting is regular~\citep{Lancaster2003}. However, regardless of the degeneracy of the eigenvalue of interest, power series expansions generally have a finite radius of convergence \citep{Fisher1999}. There is, therefore, a fundamental question that needs to be addressed: in what region of the parameter space do these power series approximations of the eigenvalues and eigenvectors converge? 

The limit in the convergence of a power series expansion is ruled by the closest singularity in parameter space, i.e., a point $\varepsilon_{\mathrm{sng}}$ such that $s(\varepsilon_\mathrm{sng})$ is singular. There may be two reasons for a singularity to exist: (i) the algebraic dependence $s(\varepsilon)$ explicitly contains a pole of the form $1/(\varepsilon-\varepsilon_\mathrm{sng})$. A notable example for compressible fluid dynamics and (thermo)acoustic problems is the dependence of the governing equations on the boundary conditions expressed in terms of an impedance $Z$, which can appear at the denominator of the governing equations, and cause a pole singularity for sound soft boundary conditions, $Z=0$;
(ii) for $\varepsilon=\varepsilon_\mathrm{sng}$ the eigenvalue problem features a defective eigenvalue with infinite sensitivity, i.e., $\varepsilon_\mathrm{sng}$ is an exceptional point (EP) in parameter space~\citep{Kato1980}. Fortunately, the closest singularity can be estimated directly from the power series coefficients, at no additional numerical cost. Using an approach that has been successfully applied in quantum mechanics~\citep{Fernandez2000}, in the following we will demonstrate how this can be achieved and how it enables us to identify the EPs closest to an eigenvalue of interest. 
%%%%%%%%%%%%%%%%%%%%%%%%%%%%%%%%%%%%%%%%%%%%
\subsection{Estimating the radius of convergence from high-order perturbation coefficients}
Close to a singularity located at $\varepsilon = \varepsilon_\mathrm{sng}$ in the parameter space, the eigenvalue parameter dependence has to be of the form
\begin{align}
 s(\varepsilon)\sim(\varepsilon-\varepsilon_\mathrm{sng})^k,
\end{align}
where $k\in \mathbb{Q}\backslash\mathbb{N}$. If $k \in \mathbb{Z}^-$, the singularity corresponds to a pole; if $k\in\mathbb{Q}^+\backslash\mathbb{N}$, the singularity corresponds to a branch-point. The value of $\varepsilon_\mathrm{sng}$ and the exponent $k$ are unknown \emph{a priori}. However, it can be shown that both quantities can be estimated from the coefficients $s_j$ of a power series that is expanded in the vicinity of (but not at) the  singularity, using the relations 
\begin{subequations}
\begin{align}
\varepsilon_\mathrm{sng}&= \varepsilon_0 + \frac{s_j s_{j-1}}{(j+1)s_{j+1}s_{j-1}-js_j^2}, \label{eq:singPos}\\
k&=\frac{(j^2-1)s_{j+1}s_{j-1}-(js_j)^2}{(j+1)s_{j+1}s_{j-1}-js_j^2}.
\label{eq:singCoeff}
\end{align}
\label{eq:singEqs}%
\end{subequations}
We refer to the perturbation techniques explained in~\citet[chapter 6]{Fernandez2000} for a detailed derivation of Eqs.~\eqref{eq:singEqs}. These estimates are asymptotic, become increasingly more accurate with the perturbation order, and converge to the closest singularity. This will be numerically shown in~\S\ref{sec:App}. When calculating the high-order sensitivity of eigenvalues around an unperturbed parameter $\varepsilon_0$, the series of eigenvalue correction coefficients will therefore converge to the actual value within a disk with radius
\begin{equation}
    R_c \equiv |\varepsilon_{\mathrm{sng}} - \varepsilon_0|,
    \label{eq:convRad}
\end{equation}
known as the radius of convergence. The value of $k$ aids in understanding the nature of the singularity: poles are identified for negative values of $k$, whereas EPs are identified by fractional values of $k$ of the form $1/a$, where $a$ is the algebraic multiplicity of the defective eigenvalue at the EP ($a=2$ for the cases considered in this article).

%%%%%%%%%%%%%%%%%%%%%%%%%%%%%%%%%%%%%%%%%%%%
\subsection{Locating EPs using perturbation theory}
\label{sec:itMet}
The closer the expansion point is to the singularity, the higher is the accuracy of the singularity estimated by Eq.~\eqref{eq:singCoeff}. This suggests a procedure that can be used to accurately locate EPs. Rather than performing a high-order expansion around $\varepsilon_0$, which becomes relatively time consuming at high orders, we adopt the following iterative scheme:
(i) calculate the expansion coefficients $s_j$ of an eigenvalue up to about order $N=10$, using the theory of~\S\ref{sec:pert};
(ii) use these coefficients to estimate the closest singularity $\varepsilon_\mathrm{sng}$ by means of Eq.~\eqref{eq:singCoeff} at the highest available order;
(iii) if the radius of convergence~\eqref{eq:convRad} is larger than a predefined threshold $\delta$, shift the expansion point to $\varepsilon_0 \leftarrow \varepsilon_0 + \varepsilon_\mathrm{sng}$ and repeat from point (i).
When $R_c < \delta$, the (shifted) expansion point coincides with the singular parameter, up to an error of order $\mathcal{O}\left(\delta\right)$. 

In general, the closest singular parameter $\varepsilon_\mathrm{sng}$ will be a complex number, even if the associated physical parameter (e.g., a time delay or a length) is a real quantity. We refer to these as non-physically realisable EPs, because one cannot perform real-world experiments with complex-valued parameters. Real-valued EPs exist, but are unlikely to be found when varying only one parameter~\citep{Seyranian2005}. A strategy to locate EPs while varying two parameters was suggested by~\cite{Orchini2020}. Even if the singularity $\varepsilon_\text{sng}$ is found in the complex plane, it nonetheless limits the convergence of the power series. This also applies when considering only real values of the parameter $\varepsilon$. Furthermore, although not realisable, the presence of complex-valued singularities has an effect on the eigenvalue trajectories. In fact, in the vicinity of EPs, eigenvalues have extremely large first order sensitivities (which become infinite at the EP). These large sensitivities cause steep variations of the eigenvalue in the complex plane, as recently analysed by~\citet{Orchini2020} and observed in, e.g.,~\citet{Bauerheim2014} and~\citet{sogaro_schmid_morgans_2019}. This phenomenon is known as mode veering~\citep{Seyranian2005}, and is the fundamental cause of the strong sensitivity of some thermoacoustic eigenvalues, and the deviation of the eigenvalue trajectories away from the first-order sensitivity predictions.

%%%%%%%%%%%%%%%%%%%%%%%%%%%%%%%%%%%%%%%%%%%%
%%%%%%%%%%%%%%%%%%%%%%%%%%%%%%%%%%%%%%%%%%%%
\section{Applications}
\label{sec:App}
We apply the methods developed in~\S\ref{sec:pert} and~\S\ref{sec:Exc} to two fundamental configurations for the investigation of thermoacoustic instabilities: a longitudinal combustor and an annular combustor. Both geometries correspond to existing experiments: respectively, the BRS combustor~\citep{Komarek2010} and the MICCA annular combustor~\citep{Bourgouin2013}. The nonlinear thermoacoustic eigenvalue problem~\eqref{eq:nep_ta} is solved for these configurations, using an $n$--$\tau$ model to reproduce the flame response at a frequency of interest.

%%%%%%%%%%%%%%%%%%%%%%%%%%%%%%%%%%%%%%%%%%%%
\subsection{Axial combustors: non-degenerate thermoacoustic modes}
\label{sec:BRS}
\begin{figure}
\centering
\includegraphics[width=.7\textwidth]{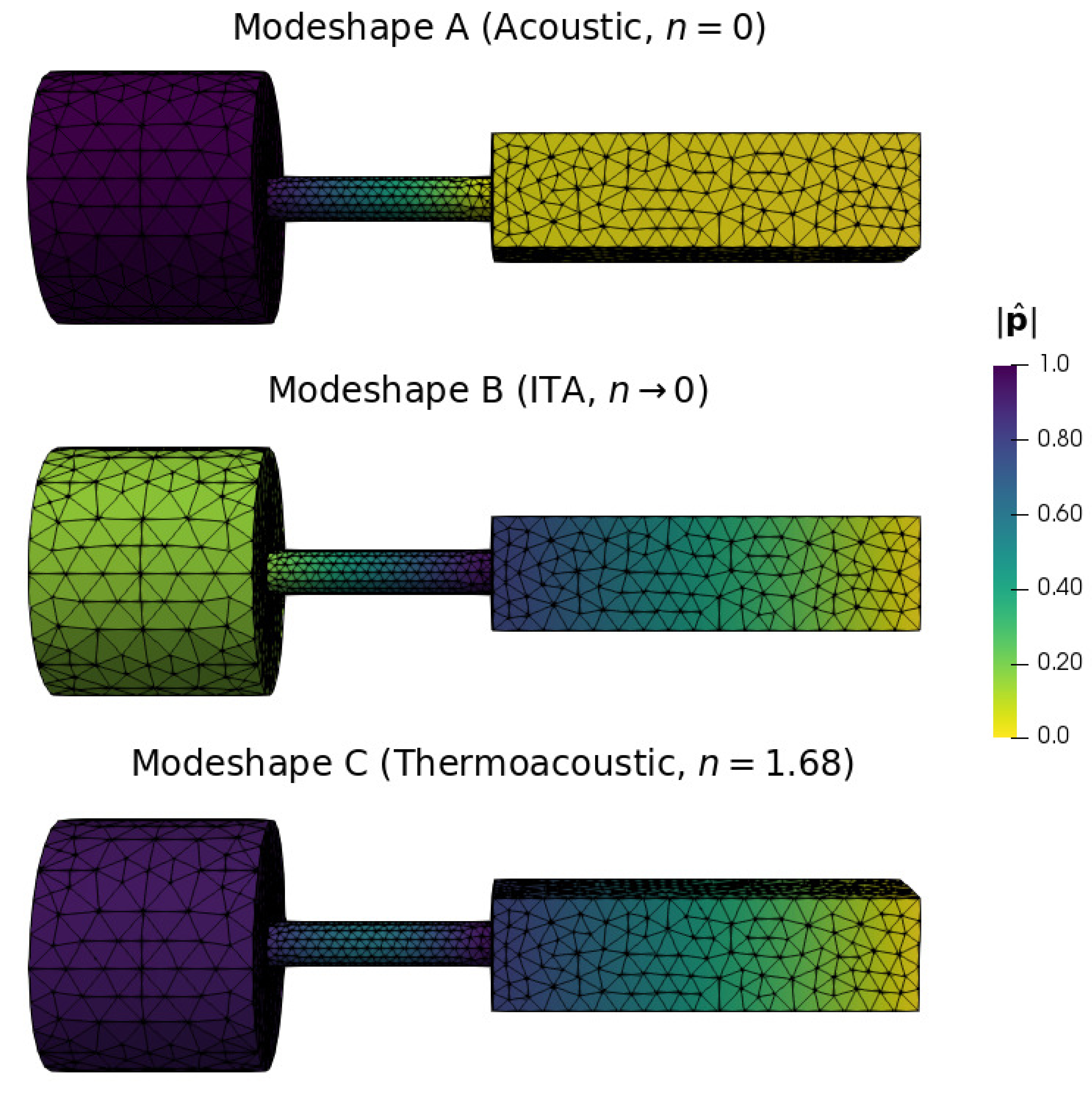}
\caption{Modeshapes of the lowest frequency acoustic mode for $\tau=4$~ms and different values of $n$. Modeshapes A and B appear for vanishing values of $n$, and correspond to an acoustic and an ITA mode, respectively. Thermoacosutic modes for finite values of $n$ will generally inherit features from both acoustic and ITA modes. This is particularly evident when the eigenvalue is close to an EP, as for the Modeshape C shown here.}
\label{fig:BRS_Modeshapes}
\end{figure}

We consider a non-degenerate thermoacoustic eigenvalue in a longitudinal combustor. In addition to demonstrating the validity of non-degenerate high-order perturbation expansions, detailed in~\cite{Mensah2020} and summarised in Eq.~\eqref{eq:nonDeg}, we will use this simpler configuration to show how the theory of~\S\ref{sec:Exc} can be used to (i) quantify the convergence limit of high-order perturbation theory and (ii) identify the closest EP to an operating condition of interest, which in turn gives essential information on the thermoacoustic spectrum and the trajectory followed by the eigenvalues when a parameter is varied. The setup we model is known as the BRS combustor; a detailed description of the geometry and the experiment is given by \cite{Komarek2010}. The 3D geomety of the model we solve is shown in Figure~\ref{fig:BRS_Modeshapes}. It consists of a cylindrical plenum, a premixing/swirling duct, and a combustion chamber with rectangular cross section.  

The BRS combustor is one of the first configurations in which thermoacoustic instabilities at a frequency that is not directly related to an acoustic mode have been experimentally observed, at about 100 Hz~\citep{Tay-Wo-Chong2012}. This instability was first generically attributed to ``flame dynamics". Later, it has been better understood and reproduced in a low-order network model by~\citet{Emmert2017}, and relabelled as intrinsic thermoacoustic (ITA) instability. ITA instabilities can be observed even in purely anechoic conditions, as they originate from the intrinsic feedback between the generation of acoustic waves by the flame and the sensitivity of the latter to upstream velocity fluctuations. %, without any influence from the specific acoustic boundary conditions. 

We discretize the thermoacoustic equation~\eqref{eq:nep_ta} on this geometry, imposing sound hard boundary conditions ($Z=\infty$) on all walls, except at the outlet, which is assumed to be sound soft ($Z=0$). A compact flame is located at the inlet of the combustion chamber. Across the flame, a temperature jump $T_2/T_1\approx5$ is imposed. The flame response is modelled with an $n$--$\tau$ model, whose values are extracted from the Flame Transfer Function (FTF) reported by~\citet{Tay-Wo-Chong2012} around the 100~Hz frequency. In particular, the time lag is assumed to be constant, as the experimentally determined FTF phase linearly decreases with frequency, with a slope $\tau\approx4$~ms. The flame gain $n$ was instead shown to be frequency dependent, with values up to 2. We choose to specify a constant value of the FTF gain, $n_0=1.68$, and consider it as a perturbation parameter. 

The thermoacoustic eigenvalue problem is solved using the open-source thermoacoustic eigenvalue solver PyHoltz~\citep{PyHoltz}. We first employ standard iterative Newton techniques to solve the eigenvalue problem. We identify a thermoacoustic mode with growth rate $\sigma=-150$~s$^{-1}$ and frequency $f=65.2$~Hz.  This eigenvalue is close to that of a Helmholtz resonant mode of the combustor, in which the plenum acts as a cavity and the premixing tube as a neck~\citep{Emmert2017}. However, its eigenvector -- Modeshape C in Figure~\ref{fig:BRS_Modeshapes} -- is not fully consistent with that of a Helmholtz mode: this mode is in fact active not only in the plenum, but also at the end of the premixing tube and at the inlet of the combustion chamber. The nature of this modeshape will be clarified in the following.

By slowly decreasing the interaction index $n$ towards zero, with steps $\Delta n=-0.02$, we track the eigenvalue trajectory in the complex plane with a continuation-like method. This eigenvalue trajectory is shown in Figure~\ref{fig:BRS_PertTeo}. We find that, in the limit $n\rightarrow0$, the growth rate of this eigenvalue tends to negative infinity, and the frequency is consistent with that of the ITA mode identified by~\citet{Emmert2017} and~\citet{Orchini2020}. Modeshape B in Figure~\ref{fig:BRS_Modeshapes} shows the magnitude of the pressure mode found when~${n=0.01}$. Its shape is indeed consistent with that of an ITA mode, as the magnitude of the eigenvector is high only around the flame~\citep{Courtine2015}.
\newsavebox{\measurebox}
\begin{figure}
\centering
\sbox{\measurebox}{%
  \begin{minipage}[b]{.49\textwidth}
  \subfloat
    []
    {\label{fig:BRS_PertTeo}\includegraphics[width=\textwidth]{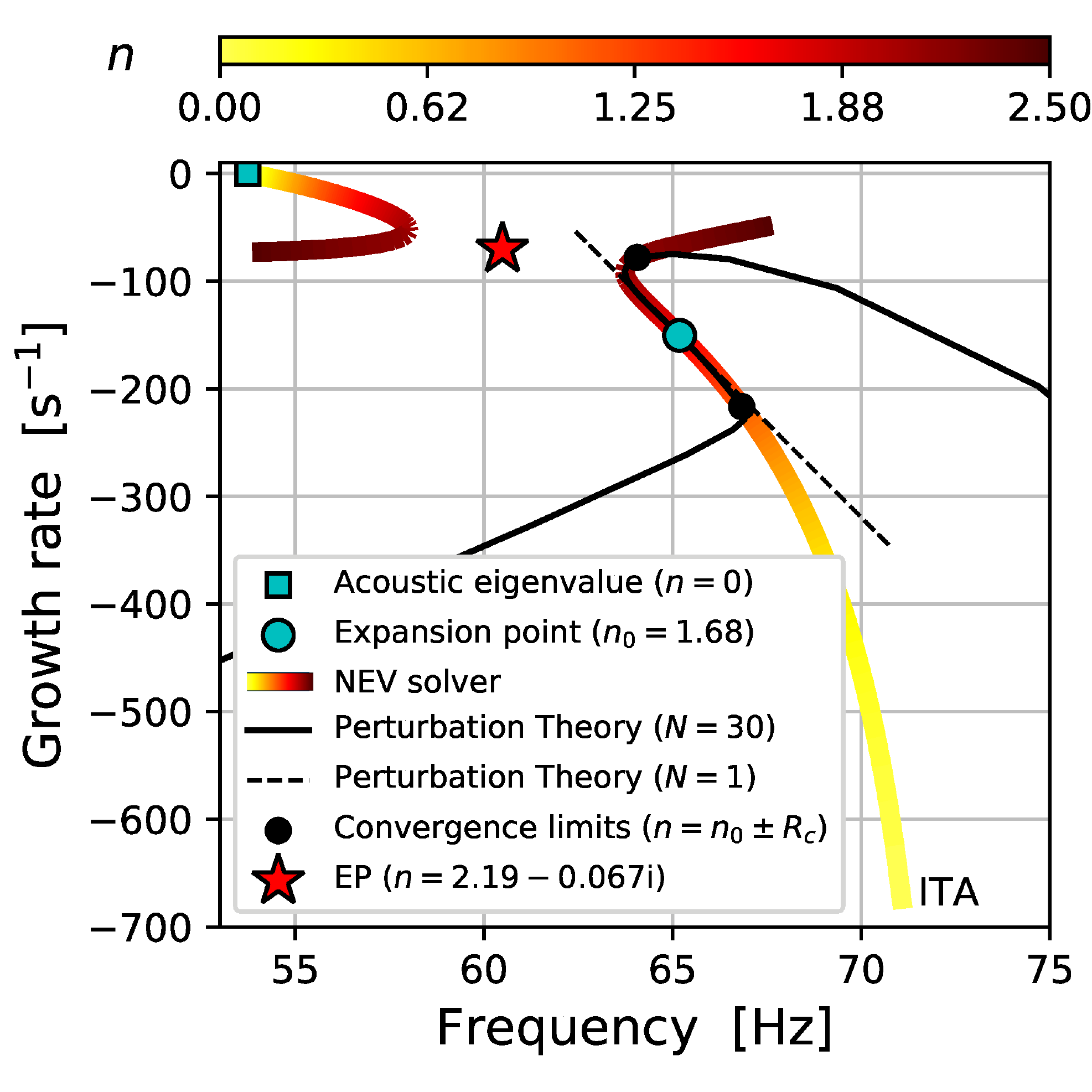}}
  \end{minipage}}
\usebox{\measurebox}
\begin{minipage}[b]{.46\textwidth}
\centering
\subfloat
  []
  {\label{fig:BRS_ConvRad}\includegraphics[trim=0mm -20mm 0mm 0mm, clip=true,width=\textwidth]{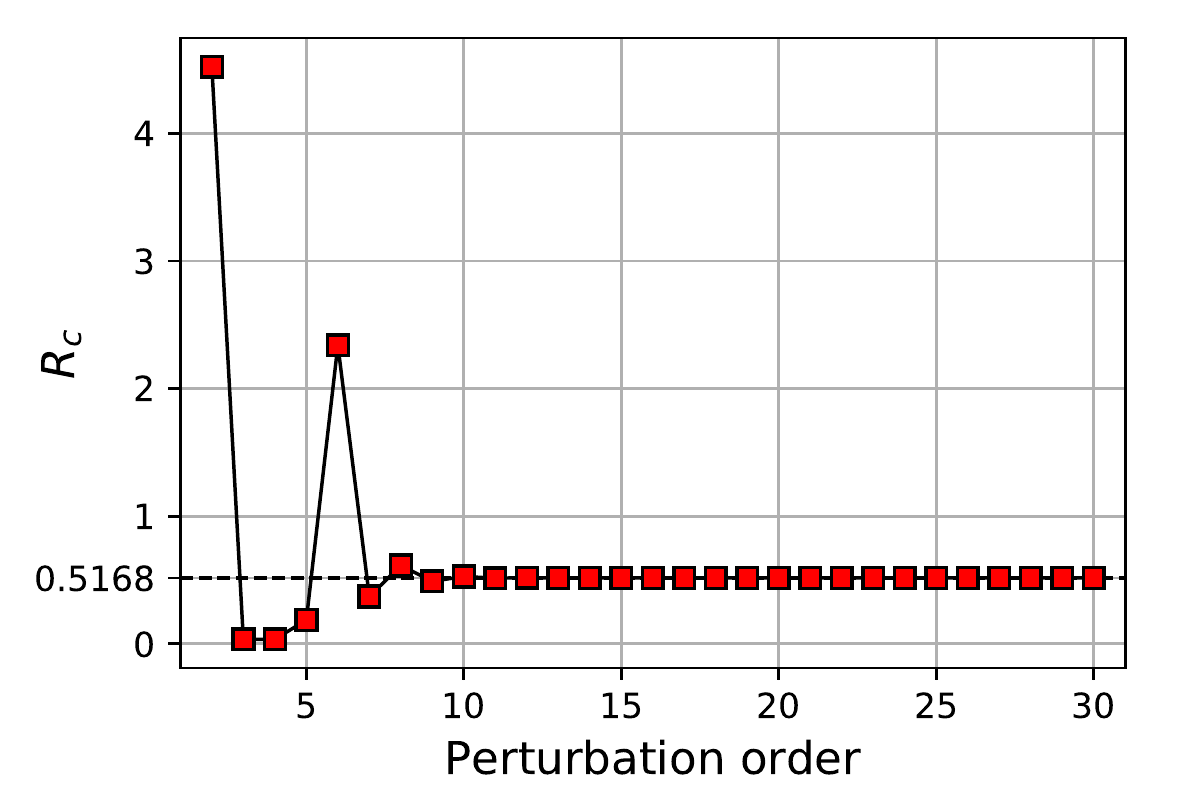}}
\end{minipage}
\caption{(a): Eigenvalue trajectories estimated with perturbation theory at 1st (dashed black line) and 30th (solid black) order, compared with exact solutions (thick shaded line). Within the radius of convergence (black markers), comparison with high-order perturbation theory is excellent. (b): Convergence of the estimated radius of convergence (and therefore location of the closest exceptional point) with the order of the perturbation expansion, from Eq.~\eqref{eq:convRad}.}
\label{fig:BRS_PertTwoFigs}
\end{figure}

We then repeat the eigenvalue tracking restarting from $n=n_0$ and increasing the interaction index $n$ up to 2.5, with steps $\Delta n=0.02$. The trajectory of the eigenvalue for large values of $n$ follows a highly nonlinear path; strong mode veering is observed. Mode veering may generally occur when a small variation in a parameter can cause two closely spaced eigenvalues to coalesce into a single degenerate eigenvalue. This degenerate eigenvalue is more likely to be an EP than a semi-simple one, unless the problem considered contains specific symmetries~\citep{Seyranian2005}, which is not the case for the thermoacoustic system considered here. We can therefore exploit perturbation theory to identify the EP responsible for the eigenvalue veering.

\begin{figure}
\raisebox{-0.5\height}{\includegraphics[width=0.55\textwidth]{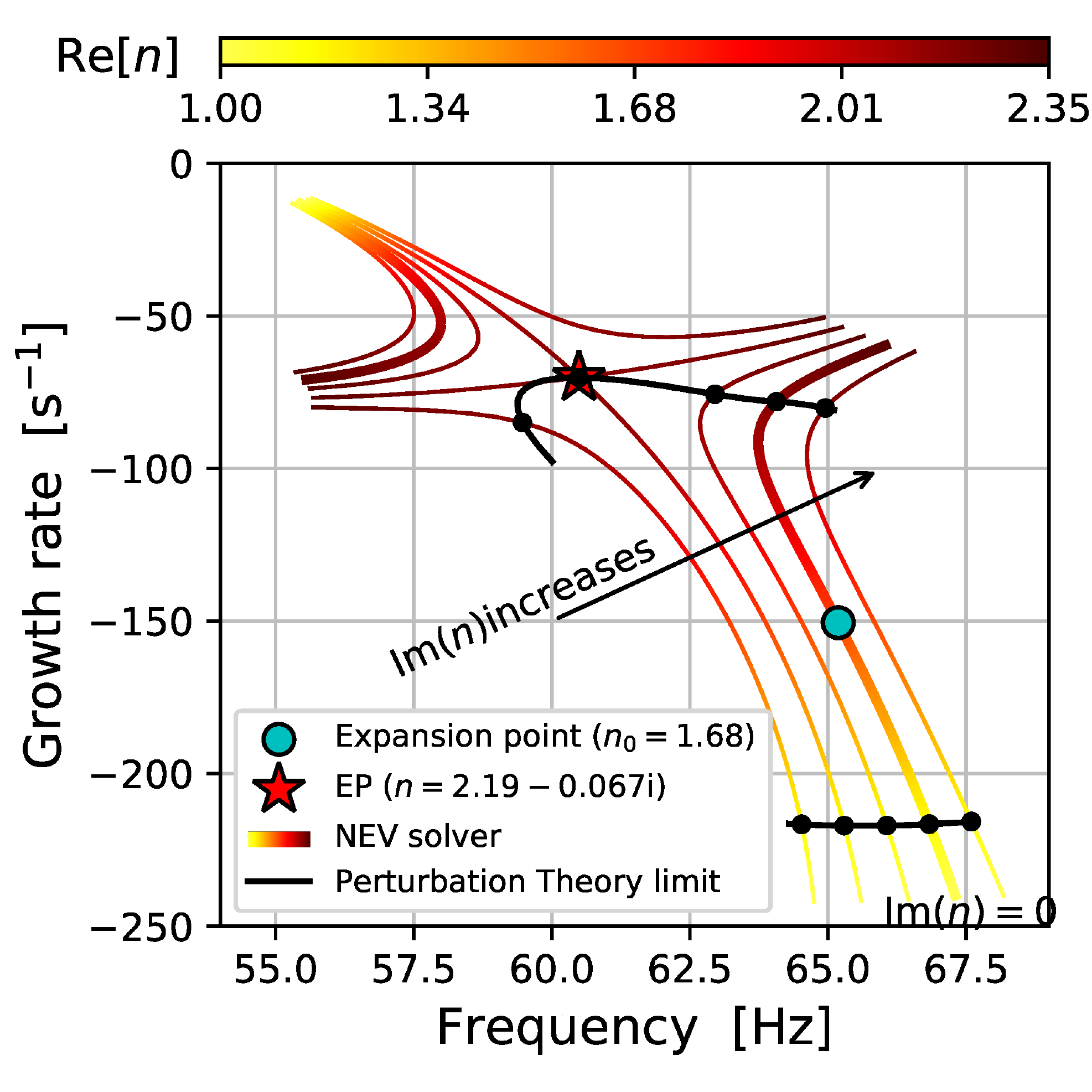}}
\raisebox{-0.5\height}{\includegraphics[width=0.4\textwidth]{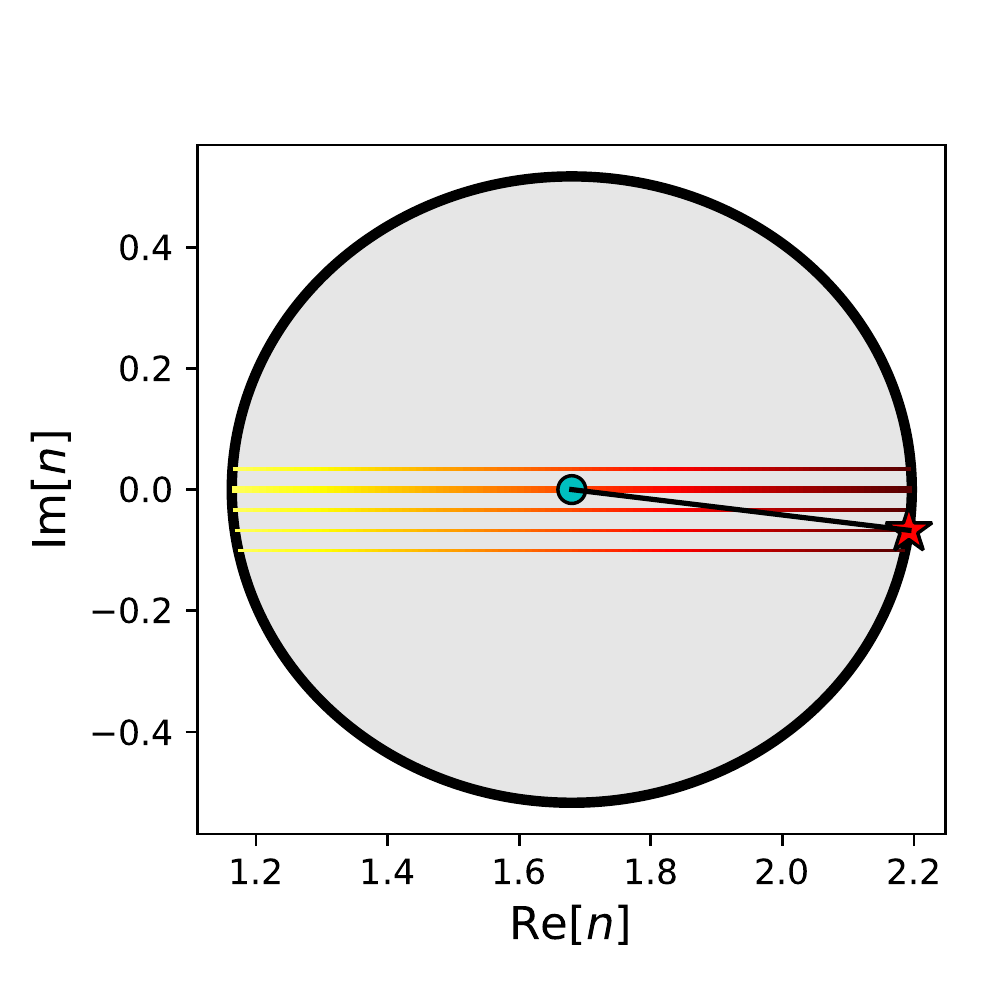}}
\caption{Left: trajectories of eigenvalues in the vicinity of the EP. The closer the trajectory is to the EP, the stronger is the veering. The thicker lines are those on which $n$ is real-valued, thus physically realisable. Right: the distance of the EP (red star) from the chosen expansion point defines the radius of convergence of the power series. The convergence region in frequency space,  between the two black lines in the left figure, is not small, which means that the perturbation method is robust.}
\label{fig:BRS_ConvRadCamilo}
\end{figure}

We consider the interaction index $n$ as a perturbation parameter, choosing as a baseline solution the value $n_0=1.68$. We apply high-order perturbation theory for non-degenerate eigenvalues, Eq.~\eqref{eq:simPert}, and calculate the coefficients of the Taylor expansion of the eigenvalue up to $30$th order. We then employ Eqs.~\eqref{eq:singEqs} to estimate the value of $n$ at which the closest singularity is found, $n_\mathrm{sng}$, and its exponent, $k$. At the highest order considered, the exponent of this singularity is $k=0.49$, close to that of a square-root branch-point. This supports the fact that the singularity is due to an exceptional point with algebraic multiplicity $a=2$, at which $k$ should have the value $1/a$. We can also estimate the radius of convergence of the power series from Eq.~\eqref{eq:convRad}. This is shown in Figure~\ref{fig:BRS_ConvRad} as a function of the perturbation order. The estimated value for $R_c$ strongly oscillates with estimates at low expansion orders, but converges to a constant value, $R_c=0.52$, at high expansion orders ($N>10$). Thus, we can determine that the Taylor expansion of the eigenvalue converges to the correct result in the entire range $n \in [1.16,2.2]$, which is a broad range for a flame gain parameter. More specifically, if we were to allow for complex values of the interaction index $n$, the eigenvalue power series expansion would converge for all $|n-n_0|<R_c$. This is shown in Figure~\ref{fig:BRS_ConvRadCamilo}. The first-order sensitivity estimate (dashed line in Figure~\ref{fig:BRS_PertTeo}) correctly predicts the slope of the eigenvalue trajectory at the expansion point, but fails in identifying the mode veering. 
The trajectory reconstructed from the expansion up to $30$th order (solid black line in Figure~\ref{fig:BRS_PertTeo}), instead, captures the veering and is almost indistinguishable from the exact solution (thick, shaded line) within the convergence limits (black markers). Outside of the radius of convergence, the expansion is not valid, and eigenvalue estimates quickly diverge.

The value of $n$ for which the eigenproblem contains an EP is $n_\mathrm{sng} = 2.19 - 0.067\mathrm{i}$, highlighted with a star in Figure~\ref{fig:BRS_ConvRadCamilo}. As this value is complex, this specific EP cannot be physically realised, although it would be possible to vary a second parameter (e.g. $\tau$) to find an EP at real-valued parameters~\citep{Mensah2018_jsv, Orchini2020}. Nonetheless, the imaginary part of $n_\mathrm{sng}$ is small: the smaller is the imaginary part of the closest EP, the stronger is the eigenvalue veering in its vicinity. The eigenvalue identified at the singular point is indicated in Figure~\ref{fig:BRS_PertTeo} with a star; the same Figure also shows how the eigenvalue trajectory exhibits the strongest veering in the vicinity of the EP. If we calculate thermoacoustic eigenvalues for real values of $n$, we can never reach exactly the EP. However, when $n=2.19$, the eigenvalue we track is very close to being defective. This implies that another eigenvalue exists in its vicinity since these two eigenvalues must coalesce at the EP. Using the contour-integration method suggested by~\citet{Buschmann2019_asme} for thermoacoustic eigenvalue problems, we, hence, search for all eigenvalues found for $n=2.19$ in the vicinity of the EP eigenfrequency. We identify two eigenvalues: one is already known, as it lies on the trajectory that was already discussed, but the second is a new eigenvalue. By applying again a continuation method, we track this newly found eigenvalue trajectory in the entire range $n\in[0,2.5]$. This trajectory is the one shown on the top-left in Figure~\ref{fig:BRS_PertTeo}. When $n=0$, this eigenvalue has zero growth rate, and corresponds to a purely acoustic mode, specifically, a Helmholtz mode of the plenum. Modeshape A of Figure~\ref{fig:BRS_Modeshapes} shows the magnitude of this mode. 

We have now all the ingredients to interpret the modeshape of the thermoacoustic mode found at $n=1.68$, Modeshape C in Figure~\ref{fig:BRS_Modeshapes}. Starting from small values of $n$, two thermoacoustic eigenvalues exist, with similar frequencies but different growth rates: the one with zero growth rate is of acoustic origin, the one with very negative growth rate is of intrinsic origin. Their eigenvectors are indicated as Modeshape A and B respectively in Figure~\ref{fig:BRS_Modeshapes}. As we increase the flame gain, the two eigenvalues first approach each other towards the EP, but eventually are repelled away from it for $n>2.19$. At the EP, not only the eigenvalues, but also the two modeshapes coalesce. The eigenvalue we considered at $n=1.68$, marked with a diamond in Figure~\ref{fig:BRS_PertTeo}, lies between the acoustic and the ITA eigenvalues, and it is relatively close to the veering region caused by the EP (Figure~\ref{fig:BRS_ConvRadCamilo}). Thus, the modeshape of the thermoacoustic eigenvalue at $n=1.68$ contains features from both the acoustic and the ITA one. This is clearly the case for Modeshape C in Figure~\ref{fig:BRS_Modeshapes}: the thermoacoustic modeshape has a strong plenum component (inherited from the acoustic modeshape) but also a strong component around the flame zone (inherited from the ITA mode). 

The case just discussed (i)  validates high-order perturbation theory for simple eigenvalues and its convergence limits, and (ii)  demonstrates that knowledge of the existence of EPs is essential for  understanding the structure of thermoacoustic modeshapes, as well as in the identification of other eigenvalues in the vicinity of trajectories that exhibit strong veering. We conclude the analysis with some remarks on the numerical cost of the calculations. On a quad-core Intel i7 processor, the Newton-like method employed for calculating eigenvalues without perturbation theory takes approximately 2~seconds to convergence for each value of the parameter $n$ considered, provided that the initial guess is reasonably accurate. Perturbation theory, on the other hand, takes approximately 30 seconds to calculate the expansion coefficients up to $30$th order, but can then be used to evaluate the eigenvalues accurately for any value of $n\in(n_0-R_c, n_0+R_c)$ at negligible computational cost.

%%%%%%%%%%%%%%%%%%%%%%%%%%%%%%%%%%%%%%%%%%%%
\subsection{Annular combustors: degenerate thermoacoustic modes}
\label{sec:deg}
We now consider an annular combustor geometry, which is directly relevant for aeronautical and power generation gas turbines. Annular combustors are known to exhibit degenerate eigenvalues~\citep{Evesque2003,Noiray2013,Bothien2015}, which arise from spatial symmetries of the system (typically discrete rotational symmetry and reflection symmetry). These degenerate eigenvalues have algebraic and geometric multiplicity two, i.e., they are semi-simple. From a physical point of view, these degenerate modes can be thought of as representing two travelling waves, one spinning in the clockwise direction and one in the counterclockwise direction, at the same frequency. A pair of degenerate thermoacoustic modes interacts nonlinearly, and a mode-selection process takes place, which can lead to the stabilisation of spinning, standing, or mixed-type thermoacoustic oscillations~\citep{Noiray2011,Ghirardo2016,Laera2017}. All these types of oscillations have been observed experimentally~\citep{Noiray2011,Worth2013a,Bourgouin2014a}. An annular combustor in which azimuthal instabilities have been investigated in detail is known as the MICCA combustor \citep{Bourgouin2014a,Prieur2017}. In addition to standing and spinning oscillations, this combustor also exhibits a more complex oscillation pattern that has been labelled slanted mode, and is believed to be due to the synchronisation of a longitudinal and an azimuthal thermoacoustic instability~\citep{Bourgouin2014,Orchini2018,Moeck2019,Yang19}. Given the interest of the thermoacoustic community in this combustor, we demonstrate the application of perturbation theory to the MICCA combustor geometry. Our goal here is not that of accurately reproducing the dynamics observed in the MICCA combustor, and we therefore simplify the linear flame response to an $n$--$\tau$ model with constant coefficients.

\begin{figure}
    \centering
    \includegraphics[width=0.5\textwidth]{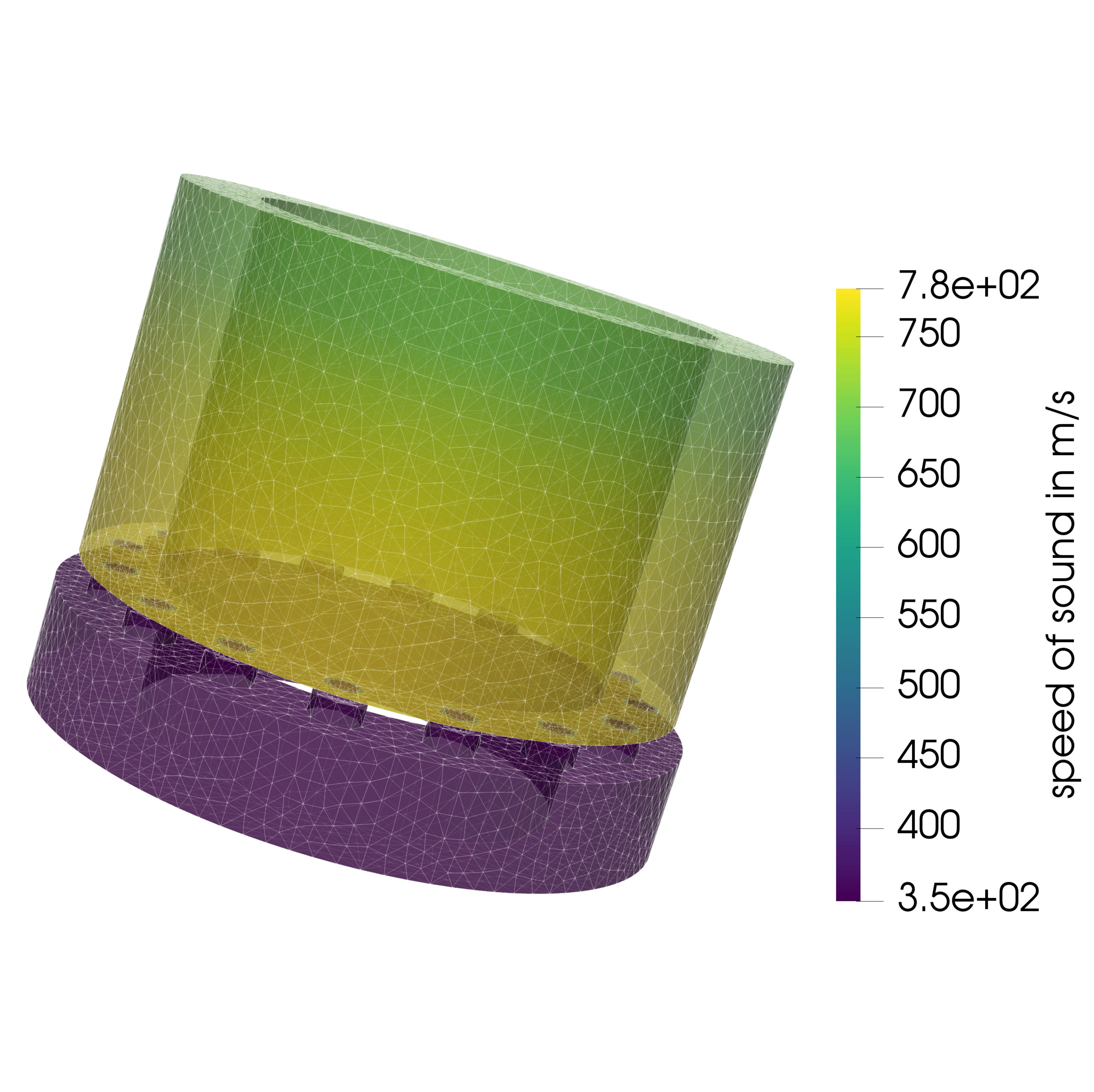}
    \caption{MICCA combustor model employed in this study. The geometrical details and speed of sound field are the same as in~\citep{Mensah2018_jtp_nonsym}. The mesh maintains the reflection symmetry of each burner segment and the discrete rotational symmetry of the whole geometry; it contains approximately 60\,000 tetrahedra.}
    \label{fig:Micca_Geo}
\end{figure}

The MICCA geometry consists of two annular sections, a plenum and a combustion chamber, connected by 16 ducts that are equispaced around the annulus. The geometry we model is shown in Figure~\ref{fig:Micca_Geo}. The geometrical details used for modelling this configuration are given by~\cite{Mensah2018_jtp_nonsym}.  We impose sound hard boundary conditions ($Z=\infty$) on all walls, and a sound soft boundary condition ($Z=0$) at the outlet of the combustion chamber. An axially varying temperature field is used (see Figure~\ref{fig:Micca_Geo}), as by~\citet{Laera2017}, with the speed of sound fixed to $c=348$~m/s in the cold plenum and ranging from $c=784$~m/s at the flame zone to $c=690$~m/s at the chamber exit. An acoustically compact flame zone is located at the exit of each duct. We fix the gain of the flame response to $n=1$, and we consider the flame time delay as the perturbation parameter, starting from the unperturbed value $\tau_0=3$~ms. Variations in the flame's time delay response are known to have a strong impact on the thermoacoustic stability~\citep{Rayleigh1878,Dowling1995}, which can be achieved, e.g., by means of fuel staging or the use of cylindrical burner outlets~\citep{Noiray2011,Krebs2002}.

We discretize the thermoacoustic equations on the MICCA geometry with finite elements, and we solve the resulting eigenproblem using PyHoltz. For the set of parameters we investigate, we identify an unstable thermoacoustic mode with frequency $f=527.78$~Hz and growth rate $\sigma=306.44$~s$^{-1}$. This eigenvalue is degenerate, with algebraic multiplicity $a=2$, and it is associated with a mode of azimuthal order 1 of the plenum cavity. This degeneracy is due to the rotation and mirror symmetries of the combustor. Thus, the eigenvalue is semi-simple, and the geometric multiplicity also equals 2. There exist therefore two linearly independent eigenvectors, spanning the 2-dimensional subspace associated with the degenerate eigenvalue, which we calculate together with their corresponding adjoint eigenvectors.

\begin{figure}
    \centering
    \includegraphics[width=\textwidth]{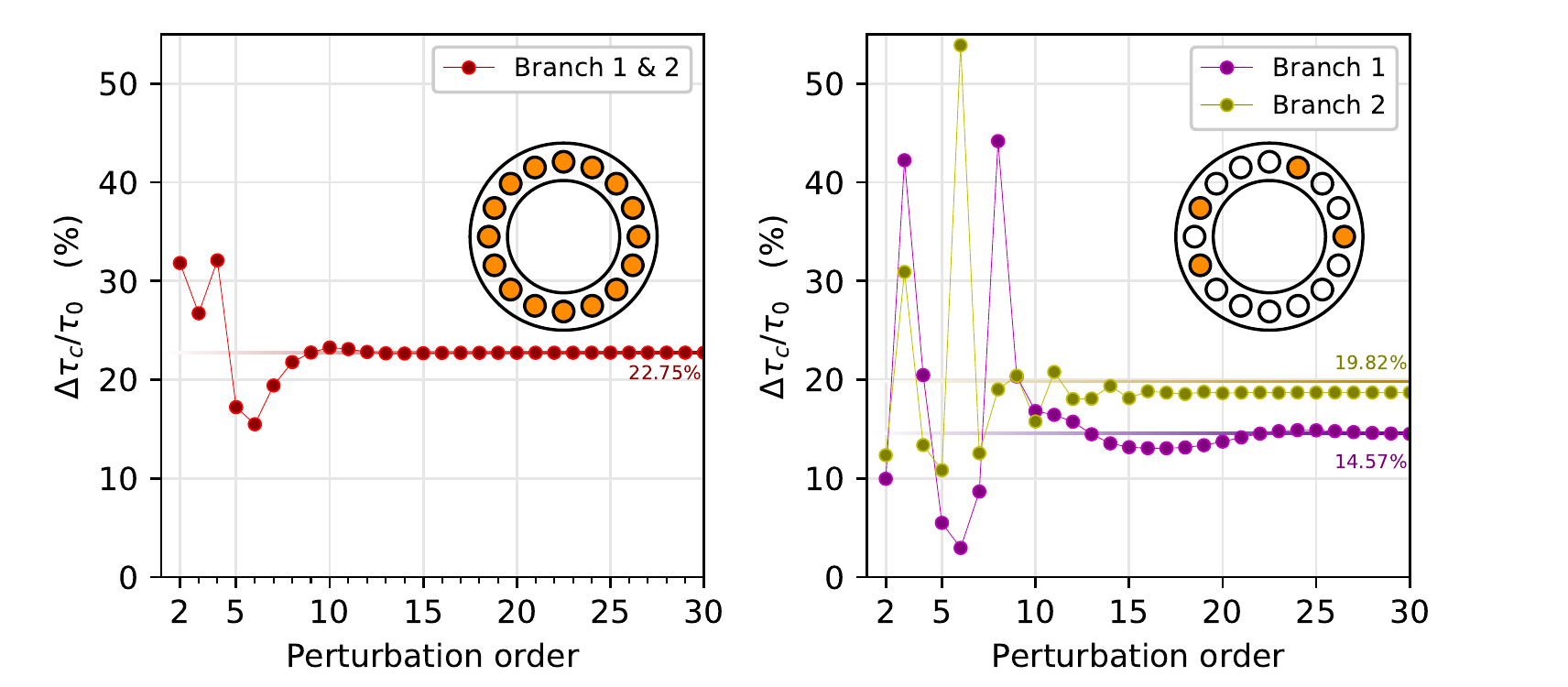}
    \caption{Radius of convergence of the perturbed eigenvalues estimated from the expansion coefficients (markers) and obtained by identifying the closest branch-point (shaded lines). The perturbation patterns are shown as insets. On the left, perturbation pattern I retains the symmetry and the eigenvalue remains degenerate regardless of the perturbation strength. On the right, perturbation pattern II breaks the geometrical symmetries, and each eigenvalue branch exhibits a different convergence behaviour.}
    \label{fig:ConvRad}
\end{figure}

We consider two types of perturbation patterns: (I) we perturb the time delay $\tau$ of all flames; (II) we perturb the time delay of certain flames only, specifically flames 1, 4, 8 and 10, counting in the counterclockwise direction~(see Figure~\ref{fig:ConvRad}). While the former pattern preserves all the symmetries of the combustor geometry, the latter pattern is chosen such that both the mirror and discrete rotational symmetries are broken. For each of the two cases, we apply degenerate perturbation theory as discussed in~\S\ref{sec:deg}.

Since pattern I preserves the symmetry, the degeneracy has not unfolded. Regardless of the magnitude of the perturbation, the eigenvalue remains degenerate with algebraic and geometric multiplicity 2. The perturbation method correctly identifies that the eigenvalue does not split since the eigenvalues of the auxiliary eigenvalue problem~\eqref{eq:auxPs} remain degenerate at all considered orders. We therefore have only one power series expansion, whose radius of convergence (shown as percentage variation from the unperturbed time delay $\tau_0$), is shown in Figure~\ref{fig:ConvRad} at various perturbation orders. The results show that high-order perturbation theory accurately predicts the evolution of the degenerate eigenvalue from the unperturbed state for variations in $\tau$ up to 22\%, a relatively large value for a flame time delay. The subspace spanned by the two linearly independent eigenvectors varies as a function of the perturbation parameter, according to Eq.~\eqref{eq:psexp}, with the coefficients of the eigenvector expansion calculated according to Eq.~\eqref{eq:Fullp}. Note that, since the problem remains degenerate, any other linear combination of the eigenvectors identified by our method would be equally valid. Our method, however, always converges to the same solution, which is constrained by the bi-orthogonalization of the perturbed eigenvectors.
\begin{figure}
    \centering
    \includegraphics[width=\textwidth]{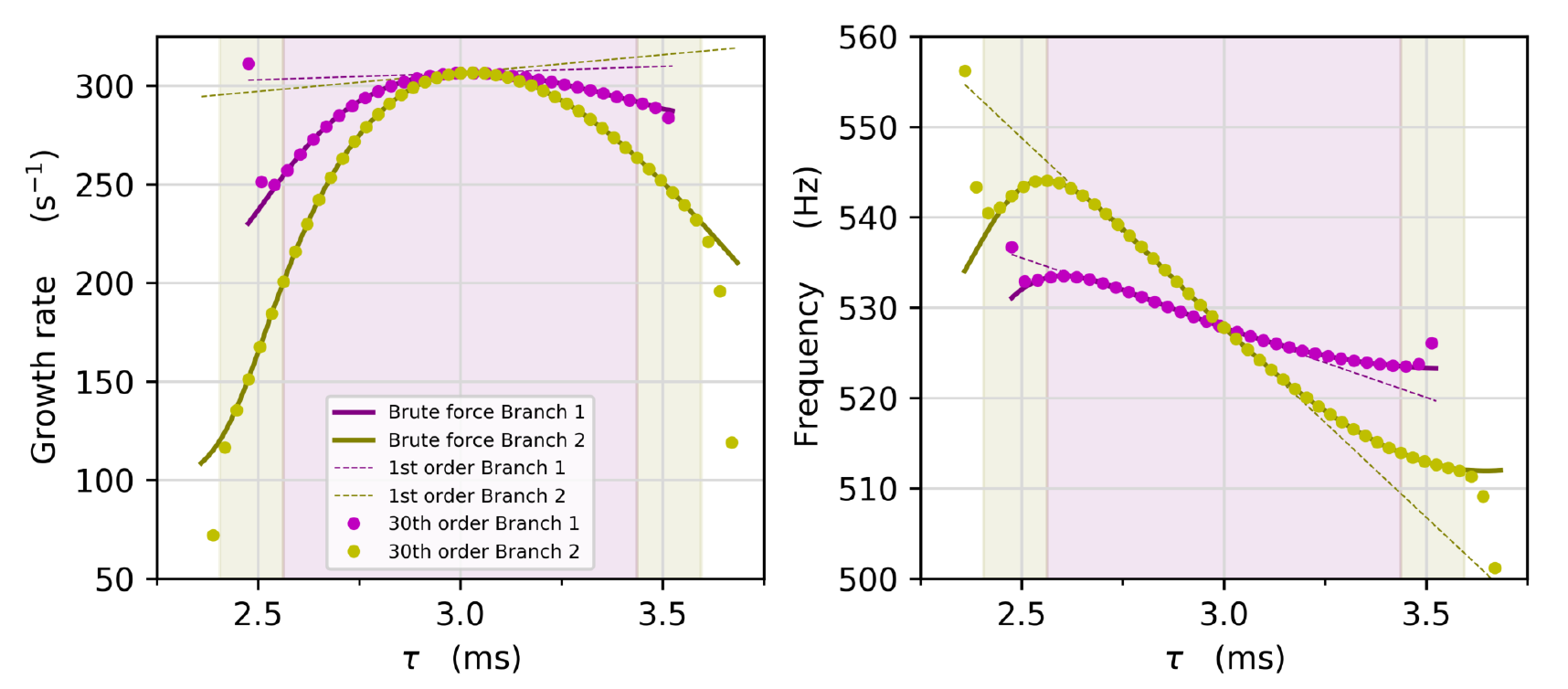}
    \caption{Comparison between the growth rate and frequency of the eigenvalues close to a degenerate state ($\tau_0=3$~ms) for the symmetry breaking perturbation pattern (II). The coloured, shaded regions indicate the radius of convergence of the respective branch. The exact solutions (solid lines) are poorly approximated by the first-order sensitivity (dashed lines) but well approximate by high-order power series expansions (markers).}
    \label{fig:Micca_Traj}
\end{figure}

On the other hand, the perturbation pattern II unfolds the degeneracy as the chosen flame staging pattern breaks the spatial symmetries. The two eigenvalues therefore follow different trajectories when varying $\tau$ (shown in Figure~\ref{fig:Micca_Traj}). We label these eigenvalue trajectories branch 1 and branch 2. Since their power series expansions have different coefficients, the estimated radius of convergence and closest EP, calculated via Eqs.~\eqref{eq:singEqs} and~\eqref{eq:convRad}, will be different for the two branches. The radius of convergence estimated from the coefficients of the two eigenvalue expansions are shown in Fig.~\eqref{fig:ConvRad}b. Note that, for branch 2, there is a small mismatch between the radius of convergence estimated via the high-order perturbation theory expansion coefficients (dotted line) and the actual distance to the closest branch-point obtained using the iteratitive procedure outlined in \S\ref{sec:itMet} (shaded line). This deviation is justified in that Eq.~\eqref{eq:singPos} is only an estimate of the location of the EP, even at high orders: the closer is the expansion point to the EP, the better is the estimate of the radius of convergence.

Despite this small mismatch, Figure~\ref{fig:ConvRad} indicates that perturbation theory should yield correct results in the prediction of the evolution of the eigenvalues in a significant range of values of $\tau$, which can vary up to 15\% and 20\% for the two eigenvalues, respectively. This is verified by comparing the reconstruction of the eigenvalue trajectories from the high-order perturbation theory with brute-force solutions (direct numerical solutions of the thermoacoustic eigenproblem~\eqref{eq:nep_ta}) at various values of $\tau\in[2.3,3.6]$~ms, shown in Figure~\ref{fig:Micca_Traj}. The shaded backgrounds in the figures indicate the range of convergence of each eigenvalue. It is evident that, within each convergence region, the brute-force and high-order perturbation results are almost indistinguishable, for both frequency and growth rate of the eigenvalues. On the other hand, as soon as perturbation theory is applied outside of the radius of convergence of its corresponding eigenvalue, the eigenvalue power series approximation rapidly diverges from the actual solution. For comparison, we have reported in Figure~\ref{fig:Micca_Traj} also the eigenvalue approximation that one would obtain by using only first-order sensitivity. For both growth rate and frequency, first-order perturbation theory correctly predicts the tangential direction along which the eigenvalues split, but fails in capturing the more complex, nonlinear behaviour of the eigenvalues at moderate changes of the perturbation parameter. This is true particularly for the growth rates shown in Figure~\ref{fig:Micca_Traj}. First-order theory predicts that increasing (decreasing) the time delay will linearly increase (decrease) the growth rate of both eigenvalues. High-order theory, instead, shows that the growth rate of the unperturbed eigenvalue is close to a maximum, so that either increasing or decreasing the flame time lag response will lead to a reduction in the growth rate of both modes.

\begin{figure}
    \centering
    \includegraphics[width=\textwidth]{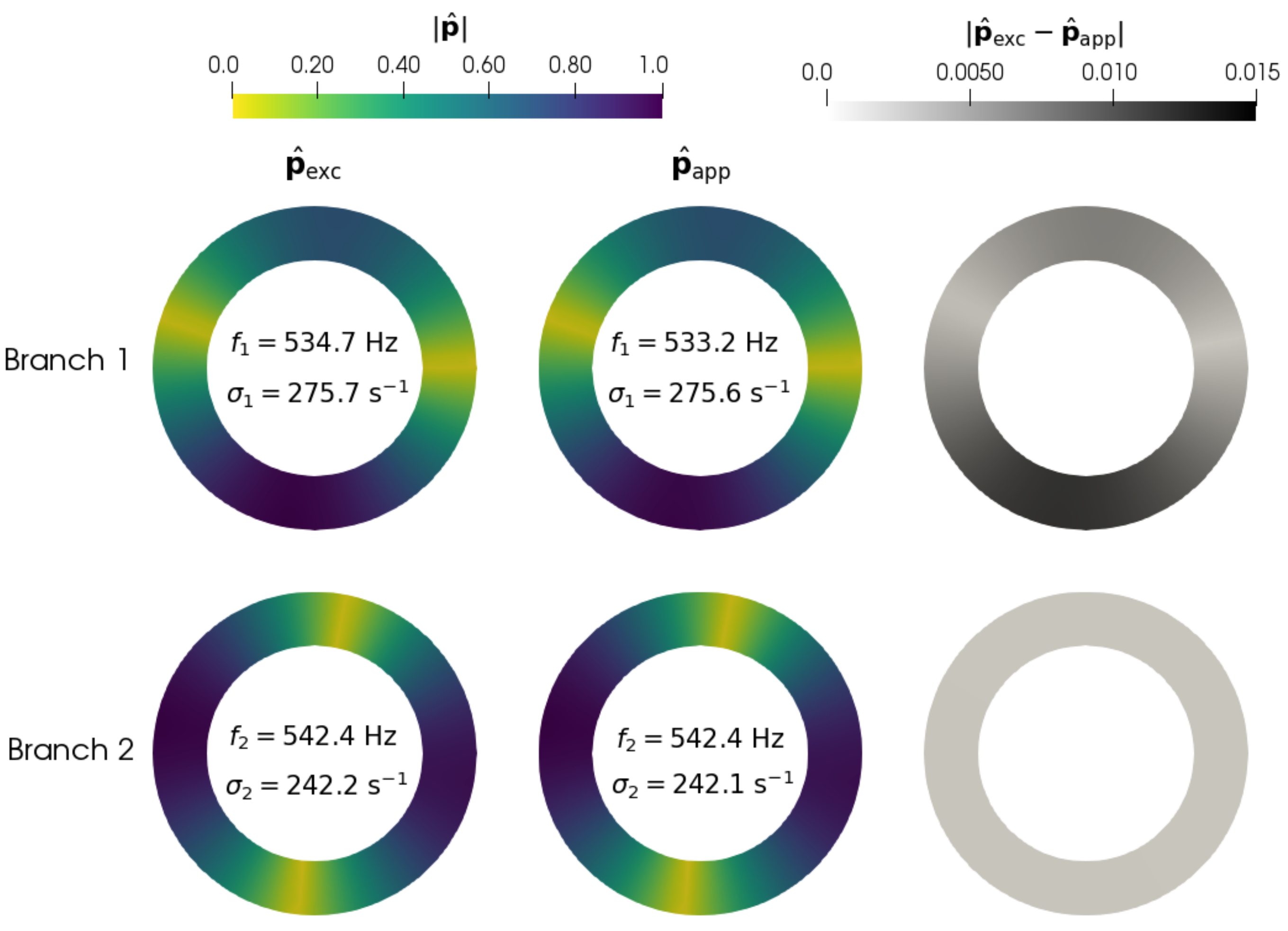}
    \caption{Cut view of the absolute value of the pressure eigenvector in the plenum of the MICCA combustor when the non-symmetric flame staging pattern II is considered. The degeneracy is resolved, and the two modeshapes that were spanning the two fold-degenerate subspace in the unperturbed case are now well-defined. The symmetries in the system's solution are lost, as correctly captured by perturbation theory. Left column: eigenvalues and eigenvectors obtained by solving the non-symmetric configuration directly with the Helmholtz solver. Middle column: eigenvalues and eigenvectors obtained from perturbation theory at 10th order, applied to the nominally symmetric, degenerate case. Right column: error between the exact and approximated eigenvectors.}
    \label{fig:Micca_Modes}
\end{figure}

Convergence results analogous to those of the eigenvalues are obtained for the eigenvectors too. Since perturbation pattern II breaks the symmetries, there exists a specific basis in the unperturbed degenerate subspace that perturbation theory must identify, which consists of the pair of eigenvectors $\hat{\bm{p}}_{0,1}$ and $\hat{\bm{p}}_{0,2}$ into which the degeneracy linearly unfolds. This is also correctly captured by the presented high-order perturbation theory, which is able to accurately reconstruct the 3D modeshape of the thermoacoustic modes at hand when varying $\tau$. As was shown in~\cite{Mensah2020}, the accuracy of the estimated eigenvectors within the radius of convergence increases at higher perturbation orders. Figure~\ref{fig:Micca_Modes} shows a plenum cut view of the magnitude of the pressure modeshapes of the split eigenvalues for $\tau=2.65$~ms. Although the asymmetry is moderate -- 4 flames have a time delay that differs by 0.35 ms from that of the other 12 flames -- the modeshapes deviate markedly from the symmetric case. This is particularly evident for the lower-frequency mode (on the top in Figure~\ref{fig:Micca_Modes}), for which the pressure nodes are visibly not aligned, and the two pressure maxima have different intensities. The modeshapes reconstructed using perturbation theory and the relative error between the exact and approximated eigenvectors are also shown. We recall that eigenvectors can always be arbitrarily scaled by a complex-valued coefficient. Thus, when comparing the exact and approximated eigenvectors, one needs to ensure that the same normalisation has been applied to the eigenvectors calculated with the exact and approximated method. This was discussed in~\citet{Mensah2020}.

We conclude the analysis with some numerical remarks. When two closely spaced eigenvalues exist, as those generated by symmetry breaking, it may be  difficult to identify both of them using standard Newton-like algorithms. This is because these iterative algorithms need an initial guess and will eventually converge to an eigenvalue solution. We observe that, often, the basin of attraction of one of these two solutions is much larger than that of the other. Unless rather accurate initial guesses are provided to the iterative algorithm, only one of the two solutions will be identified. High-order perturbation methods are useful also in this respect. Within the radius of convergence, the eigenvalue estimates they provide are suitable initial guesses for the actual eigenvalues of the system, as shown above, and Newton-like methods are able to identify both closely spaced eigenvalues in a few iterations. The brute-force results shown in Figure~\ref{fig:Micca_Traj} have been calculated using this effective strategy. An alternative is to use Beyn's contour-integration method, which is able to identify all eigenvalues inside a circle in the complex-frequency space~\citep{Beyn2012,Buschmann2019_asme}. Both methods have their pros and cons. High-order perturbation methods require only one longer calculation, needed to obtain the eigenvalue expansion coefficients, and then only short calculations to (i) obtain good initial guesses and (ii) converge to the eigenvalue(s) of interest via Newton methods, for any value of the perturbation parameter within the radius of convergence. Beyn's method, instead, requires a moderately long calculation for each of the perturbation parameter values one is interested in, but does not suffer from radius of convergence limitations -- although for very large perturbations, the split eigenvalues may be far from each other, which requires the integration over a large circle, with increasing computational effort needed. We are of the opinion that no general conclusion can be made regarding which of the two methods is preferable. The trade-off depends on the size of the eigenproblem at hand and on the range parameters one wants to investigate.

%%%%%%%%%%%%%%%%%%%%%%%%%%%%%%%%%%%%%%%%%%%%
%%%%%%%%%%%%%%%%%%%%%%%%%%%%%%%%%%%%%%%%%%%%
\section{Expansion of defective eigenvalues at exceptional points}
\label{sec:defective}
At defective points, the theory described in~\S\ref{sec:pert} breaks down. This is because the eigenvectors of a defective eigenvalue lose the bi-orthogonality property~\eqref{eq:ortoBio}. We have demonstrated the existence of EPs, which are defective eigenvalues (\S\ref{sec:Exc}), in the spectrum of thermoacoustic systems. To make our technique applicable to all types of eigenvalues, in this section, we present the perturbation theory at defective eigenvalues. 
For defective eigenvalues, the following relation between direct and adjoint eigenvectors holds in place of~\eqref{eq:ortoBio}:
\begin{equation}
\inner{\hatt{\bm{p}}^\dagger_{\mathrm{def}}}{\mathsfbi{L}_{1,0} \hatt{\bm{p}}_\mathrm{def}} = 0.
\label{eq:ortoBioDef}
\end{equation}
This is known as self-orthogonality in non-Hermitian quantum mechanics~\citep{Moiseyev2011,Heiss12} and invalidates the derivation of the expansion equations of~\S\ref{sec:solvCond}. The latter relies on expressions of the form analogous to those shown in Eq.~\eqref{eq:nonDeg}, in which the scalar product in the denominator vanishes in the defective eigenvalue scenario due to the self-orthogonality condition~\eqref{eq:ortoBioDef}. Consequently, already the first-order sensitivity of the eigenvalues diverges to infinity, and it is not possible to expand the dependence of the eigenvalues on a parameter into a power series. 

However, eigenvalues at an EP can be expanded into a fractional power series, also known as Puiseux series. In particular, for a defective eigenvalue with algebraic multiplicity $a$, it is possible (under mild assumptions) to expand the eigenvalue as follows~\citep{Leung1990,Lancaster2003}:
\begin{equation}
    s(\varepsilon) \approx s_0 + \sum_{j=1}^N s_j \varepsilon^{j/a}.
    \label{eq:Puiseux}
\end{equation}
By using this ansatz, and introducing the concept of generalised eigenvectors for defective eigenvalues (see Supplementary Material~\S\ref{sec:Supp1}), it is possible to follow the steps of~\S\ref{sec:solvCond} to derive arbitrary high-order equations for the calculation of the coefficients of the Puiseux series~\eqref{eq:Puiseux}. It is not the aim of this contribution to provide the entire (and lengthy) derivation of these expressions. Our goal is to show that expansions at EPs are possible, and that the Puiseux coefficients can be evaluated by means of adjoint methods. Additional details on the derivation of the Puiseux expansions at EPs are provided in the Supplementary Material.

Here, we shall focus only on the special case of defective eigenvalues with algebraic multiplicity $a=2$ and (consequently) geometric multiplicity 1. In this case, it can be shown (see Eq.~\eqref{eq:fredPuis1} in the Supplementary Material) that the first coefficient of the Puiseux series expansion is given by
\begin{equation}
    s_1 = \pm \sqrt{-\frac{\inner{\hat{\bm{p}}^\dagger_{\mathrm{def}}}{\mathsfbi{L}_{0,1} \hat{\bm{p}}_\mathrm{def}}}{\inner{\hat{\bm{p}}^\dagger_{\mathrm{def}}}{\mathsfbi{L}_{1,0} \hatt{\bm{p}}_\mathrm{gen}}+\inner{\hat{\bm{p}}^\dagger_{\mathrm{def}}}{\mathsfbi{L}_{2,0} \hat{\bm{p}}_\mathrm{def}}}},
    \label{eq:defSens}
\end{equation}
where $\hatt{\bm{p}}_\mathrm{gen}$ is the generalised eigenvector, defined by $\mathsfbi{L}_{0,0}\hatt{\bm{p}}_\mathrm{gen} \equiv -\mathsfbi{L}_{1,0}\hat{\bm{p}}_{\mathrm{def}}$. Note the differences between~\eqref{eq:defSens} and the first-order ordinary sensitivity equation, which reads $s_1 = - \inner{\hatt{\bm{p}}^\dagger_0}{\mathsfbi{L}_{0,1} \hatt{\bm{p}}_0}/\inner{\hatt{\bm{p}}^\dagger_0}{\mathsfbi{L}_{1,0} \hatt{\bm{p}}_0}$. First, two branches (the + and - roots) stem from the sensitivity equation for defective eigenvalues~\eqref{eq:defSens} due to the $a=2$ algebraic multiplicity. Second, a square root appears in the equation, highlighting the fact that the defective eigenvalue is a branch-point. This is due to the fact that the first coefficient in the Puiseux series expansion is determined at second order (whereas first-order power series expansion coefficients are determined at first order). Lastly, a new term appears in the denominator, which involves the second derivative of the operator $\mathsfbi{L}$ with respect to the eigenvalue. This results from the fact that the defective eigenvalue sensitivity is determined at second order and that bi-orthogonality conditions do not hold.

The above sensitivity equation represents an exception to the standard sensitivity equations for semi-simple eigenvalues outlined in~\S\ref{sec:solvCond}, and should be used if and only if the eigenvalue of interest is a defective eigenvalue. Note that, in practice, it is unlikely to converge exactly to a defective eigenvalue with numerical methods. Both the series~\eqref{eq:singPos} -- which converges to the closest EP in the complex-parameter space -- and the method outlined in~\citet{Orchini2020} -- which converges to an EP in the real parameter space -- are iterative numerical methods, whose accuracy is limited by machine precision. Due to the infinite sensitivity of eigenvalues at EPs, even a very small deviation in the estimation of the defective eigenvalue is sufficient to strongly affect the eigenvalue sensitivity, causing large errors in the results of Eq.~\eqref{eq:defSens}. The latter should therefore be applied only to problems for which EPs can be identified analytically. 

\subsection{Application to a 1D thermoacoustic model}
We shall now provide an application of Puiseux series expansion to numerically show that the sensitivity to small perturbations of eigenvalues at EPs is not polynomial, but scales with powers of $\varepsilon^{1/2}$. The model we use is a single-mode Galerkin expansion of the thermoacoustic equations in an acoustically open tube~\citep{Juniper2018}, expressed in non-dimensional units, for which the scalar operator $L$ is given by
\begin{equation}
    L(s) \equiv s^2 + 2 \pi \beta e^{-s\tau} + \pi^2.
    \label{eq:1Galerk}
\end{equation}
We wish to find defective eigenvalues of the above equation, which are obtained when
\begin{equation}
    L(s) = 0 \quad \mbox{and} \quad \frac{\partial L}{\partial s} \equiv L_{1,0} = 0.
    \label{eq:defConds}
\end{equation}
These are generally only necessary conditions for a defective point, as they would also be satisfied for a semi-simple degenerate eigenvalue with algebraic multiplicity of (at least) two. However, as shown in~\citet{Seyranian2005}, when degenerate eigenvalues arise the occurrence of a defective eigenvalue is almost certain in a system without symmetries -- semi-simple degenerate eigenvalues exist, but have a negligible measure compared to that of the defective ones. No symmetries are present here, so Eqs.~\eqref{eq:defConds} effectively identify EPs.

From Eqs.~\eqref{eq:1Galerk} and~\eqref{eq:defConds} it can be shown that choosing $\beta \tau e = \pm 1$ for $\pi\tau > 1$, and with the additional constraints that
$\tan{[\left(\pi\tau\right)^2-1]^{1/2}}=[\left(\pi\tau\right)^2-1]^{1/2}$, yields a defective eigenvalue with algebraic multiplicity 2 and geometric multiplicity 1 of the form
\begin{equation}
    s_{\mathrm{def}} = -\frac{1}{\tau} \pm \mathrm{i} \frac{\sqrt{\left(\pi\tau\right)^2-1}}{\tau}.
\end{equation}
The non-dimensional parameters at which we identify an EP are $\tau=1.4653$ and ${\beta=0.2511}$; the corresponding defective eigenvalue is $s_\mathrm{def}=-0.6825 + 3.0666\mathrm{i}$.

Let us now consider a perturbation expansion of the defective eigenvalue around $\tau$. Since we have imposed that $L_{1,0} = 0$, it is not correct to use Eq.~\eqref{eq:nonDeg} to try to identify polynomial coefficients, as the latter equation diverges. Instead, we can calculate the first coefficient of the Puiseux series using Eq.~\eqref{eq:defSens}, which yields the following expansion for the eigenvalues around the EP:
\begin{equation}
    s = s_{\mathrm{def}} \pm \sqrt{\frac{2\pi\beta s_{\mathrm{def}} e^{-s_{\mathrm{def}}\tau}}{1 + \pi\beta\tau^2 e^{-s_{\mathrm{def}}\tau}}} (\Delta\tau)^{1/2} + \mathcal{O}\left(\Delta\tau\right).
    \label{eq:Puis1}
\end{equation}
Because this example is 1-dimensional, the direct and adjoint eigenvectors associated with the EP equal 1, and the generalised eigenvector vanishes.

\begin{figure}
    \centering
    \includegraphics[width=0.48\textwidth]{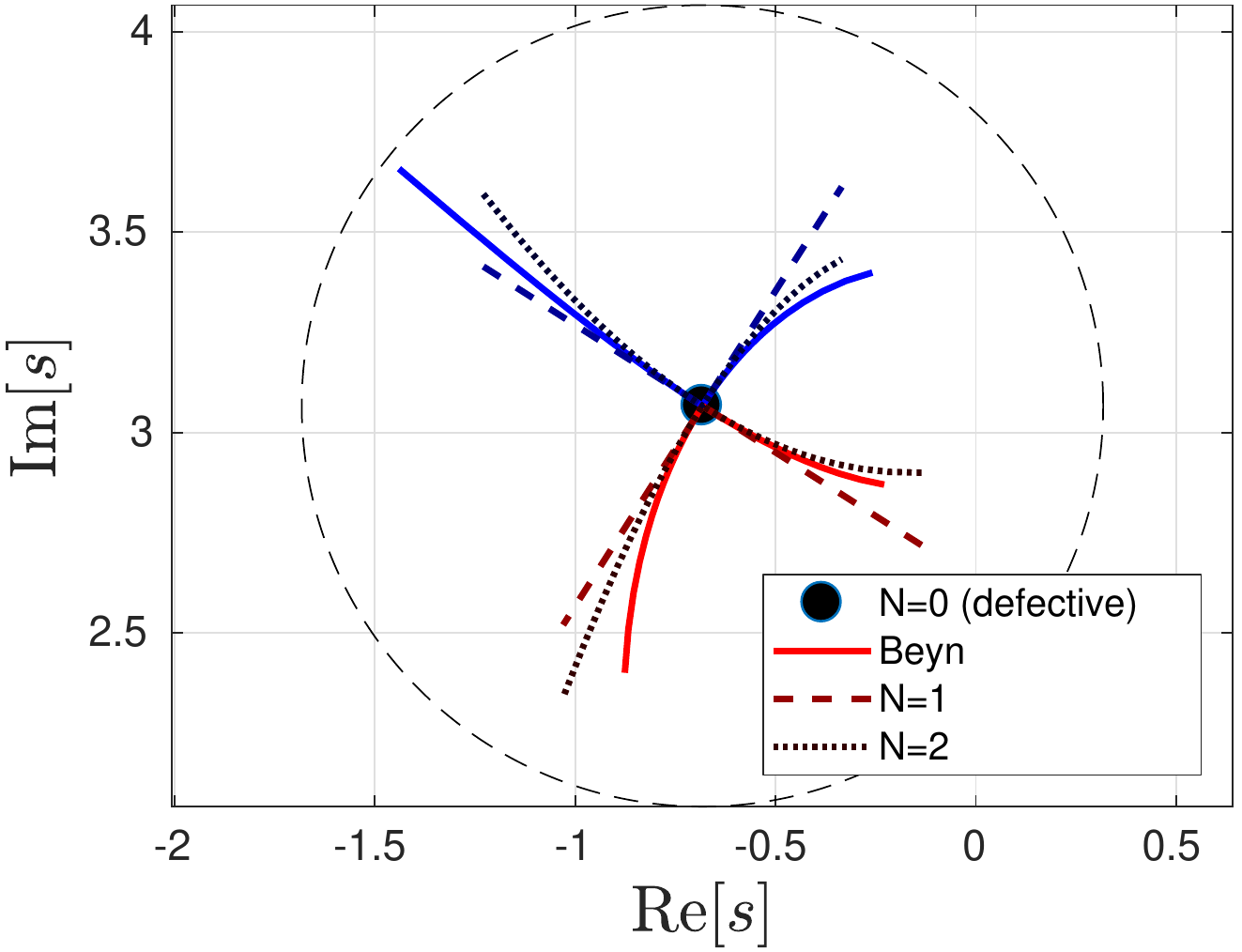}
    \includegraphics[width=0.48\textwidth]{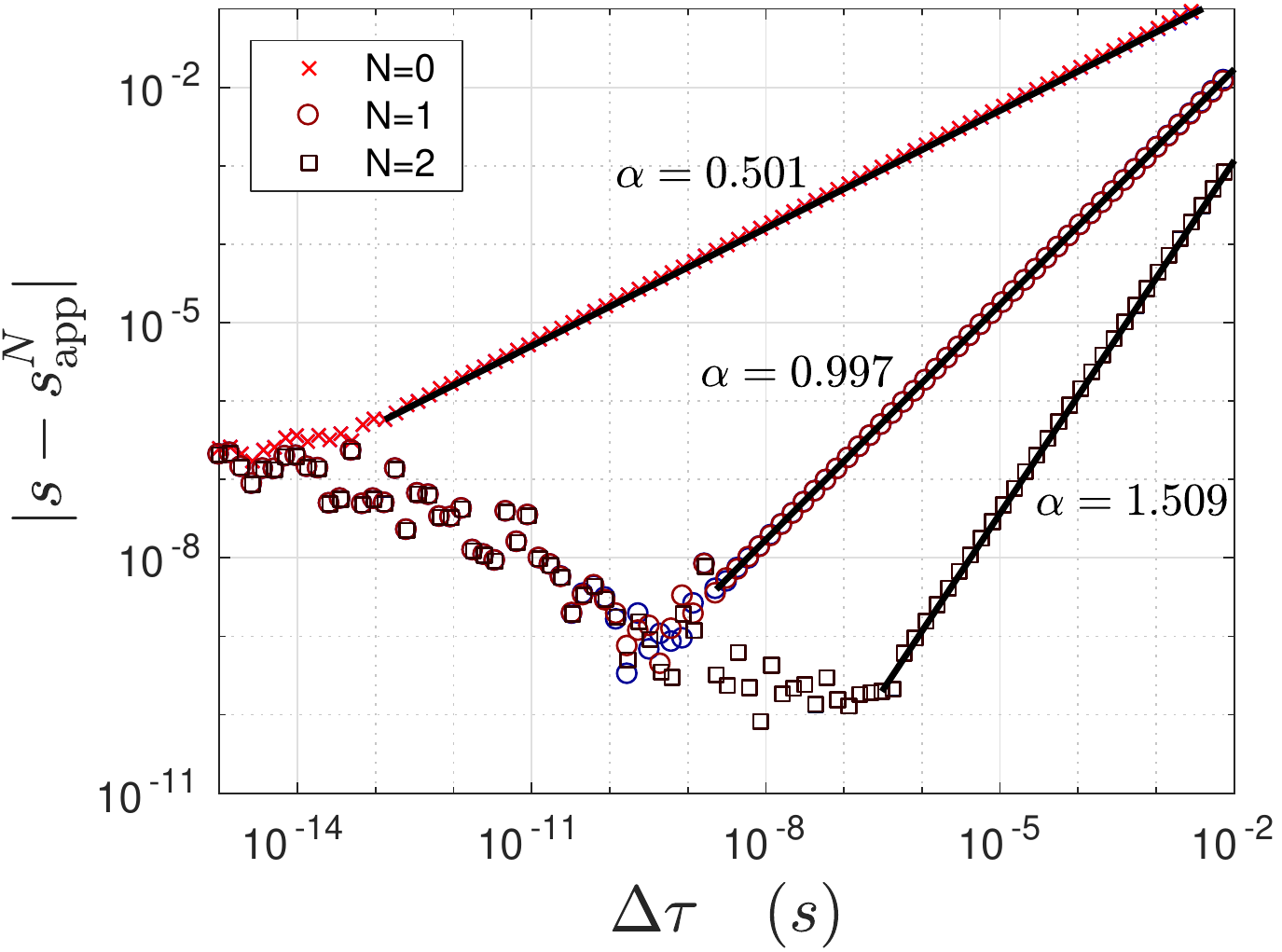}
    \caption{Left: trajectories of two eigenvalues coalescing at an EP (solid lines), and 1st (dashed) and 2nd (dotted) order Puiseux approximations of these trajectories. The discontinuous branch-point behaviour at the EP is correctly captured by the Puiseux approximations. Right: relative error between the exact eigenvalues and their approximations by Puiseux series, at various orders in log-log scale. The error scales as $(\Delta\tau)^{1/2}$, which is consistent at an EP with algebraic multiplicity 2.}
    \label{fig:DefPoint}
\end{figure}

Figure~\ref{fig:DefPoint}a shows the evolution of the eigenvalues around the EP while varying $\tau$. The solid lines are obtained by repeatedly applying Beyn's contour-integral method while varying the flame's time delay. The colour indicates the evolution of each tracked eigenvalue: at the EP (black dot) the eigenvalues coalesce, and the trajectories have a discontinuous first-order derivative, with cusp angles of 90 degrees. The eigenvalue trajectories approximated with the first-order Puiseux series~\eqref{eq:Puis1} are shown with dashed lines. The Puiseux series expansions accurately capture the angle along which the two cusps formed by the eigenvalue trajectories intersect at the EP, and are locally tangent to the exact trajectories. Figure~\ref{fig:DefPoint}b shows the relative error between the actual eigenvalue and that estimated by the Puiseux series for the red branch only; analogous results are obtained for the other branch. At expansion order $N=0$, the error equals to the distance between the eigenvalue for a given $\tau$ and the eigenvalue found at the EP. For very small variations of $\tau$, machine precision limits the error to a lower bound. However, when increasing the variation in the time delay, $\Delta \tau$, the relative error follows a linear trend in log-log scale. Linear regression identifies the slope of this line to be $\alpha=0.501$, implying that the error scales with $\Delta\tau^{1/2}$. This is the leading term in the Puiseux series~\eqref{eq:Puiseux} for algebraic multiplicity $a=2$. When calculating the error including the first term of the Puiseux expansion, instead, the relative error generally decreases and its slope becomes $\alpha\approx 1$, which is the power of the leading error in Eq.~\eqref{eq:Puis1}. A further increase in accuracy by a factor of $\Delta\tau^{1/2}$ is observed when including also the second term of the Puiseux series, $N=2$, whose expression for the model investigated here is provided in Eq.~\eqref{eq:2ndPuis} of the Supplementary Material. The trajectories of the eigenvalues reconstructed including also the second term of the Puiseux series are shown in Figure~\ref{fig:DefPoint}a with dotted lines. They provide better approximations of the actual eigenvalues for larger deviations from the EP, and they are able to capture both the local angle and also the curvature of the cusps that intersect at the EP.

%%%%%%%%%%%%%%%%%%%%%%%%%%%%%%%%%%%%%%%%%%%%
%%%%%%%%%%%%%%%%%%%%%%%%%%%%%%%%%%%%%%%%%%%%
\section{Conclusions}
\label{sec:Conclusions}
In this study, we have established interconnections between several topics that are essential for the investigation of thermoacoustic stability, namely symmetry-induced degeneracies, high-order adjoint perturbation theory, the origin of thermoacoustic modes, and exceptional points. We have first extended adjoint-based high-order perturbation formulae for thermoacoustic eigenproblems, currently available only for simple eigenvalues, to semi-simple degenerate eigenvalues, typical of annular configurations. Although degenerate perturbation theory could be extended to arbitrary level of degeneracies, the 2-fold degenerate case, which arises due to spatial symmetries of combustor configurations, is the most relevant for applications in thermoacoustics, and is the one that has been considered in this study.
%We have shown that, for degenerate eigenvalues, an auxiliary eigenvalue problem needs to be solved, to determine whether the considered perturbation breaks or retains the symmetry, and therefore the eigenvalue degeneracy. 

%We have discussed how w
Well-defined boundaries of validity of eigenvalues reconstructed with perturbation methods can be defined by evaluating the radius of convergence of the power series, directly from their coefficients. The radius of convergence is generally finite but not necessarily small. In fact, it can be sufficiently large to use perturbation theory for investigating the effects of parametric variations on the thermoacoustic spectrum in a broad parameter range of physical relevance. The existence of a finite limit of convergence is due to singularities in the thermoacoustic eigenproblems that arise at exceptional points. Exceptional points appear in the spectrum of thermoacoustic systems when eigenvalues of acoustic and/or ITA origin coalesce, and we have shown how they can be identified by means of the coefficients of high-order perturbation methods. At exceptional points, eigenvalues have infinite sensitivities; moreover, the eigenvalue trajectories that form when a parameter is varied exhibit strong veering in the vicinity of an EP. This explains the high sensitivity of some thermoacoustic modes that has recently been discussed in the literature.

We applied the presented theory to two thermoacoustic configurations that model existing experiments. We demonstrated that the identification of exceptional points facilitates the prediction and understanding of the trajectories and modeshapes of thermoacoustic eigenvalues as the system parameters are varied. We have shown that degenerate perturbation theory is capable of correctly predicting whether a given perturbation pattern breaks or retains the degeneracy of semi-simple eigenvalues found in annular combustors. When splitting occurs, the method correctly predicts the two trajectories that are followed by both eigenvalues, as well as the eigenvector basis into which the unperturbed problem unfolds. All these results are valid within a radius of convergence that was accurately calculated from the power series coefficients. The framework enables one to predict the strong veering that thermoacoustic eigenvalues may experience, and explain the latter via the identification of the closest EP.

Lastly, we have shown how adjoint-based methods allow also for the calculation of the coefficients of Puiseux series, which can be used to expand the eigenvalues found at exceptional points. Combined together, our findings provide a comprehensive and efficient theory that can be applied to investigate the thermoacoustic spectrum in the vicinity of any eigenvalue (simple, semi-simple or defective) while varying a parameter.

Our results show that perturbation theory is a powerful tool that enables the efficient and accurate assessment of the effects that parametric variations have on the thermoacoustic spectrum. Its results are valid in a finite but broad range of design parameters, and can be exploited in the execution of tasks relevant to the thermoacoustic community such as design optimisation and uncertainty quantification. In fact, although the trajectories followed by thermoacoustic eigenvalues are nonlinear, in particular in the vicinity of an EP, often only the linear first-order sensitivity is considered for the prediction of eigenvalue variations. This can lead to a misinterpretation of the behaviour of the thermoacoustic spectrum. The high-order adjoint-based method proposed here constitutes a significant step forward, as it enables the accurate estimation of the nonlinear behaviour of thermoacoustic eigenvalues in a wide parametric range. This helps in understanding and predicting the effect that real-world finite perturbations have on thermoacoustic stability. It can also be leveraged in advanced optimisation methods, by providing nonlinear descent information that can lead to a quicker identification of local optima.
%%%%%%%%%%%%%%%%%%%%%%%%%%%%%%%%%%%%%%%%%%%%
%%%%%%%%%%%%%%%%%%%%%%%%%%%%%%%%%%%%%%%%%%%%
\section*{Acknowledgment}
A. Orchini acknowledges support from the Alexander von Humboldt Foundation and the German Research Foundation (DFG Project Nr.\ 422037803). L. Magri acknowledges support from the Royal Academy of Engineering Research Fellowship Scheme, and support from the Technical University of Munich - Institute for Advanced Study, funded by the German Excellence Initiative and the European Union Seventh Framework Programme under grant agreement no. 291763.

\noindent
The authors report no conflict of interest.

\appendix
\section{Explicit expressions of perturbation equations}
\label{app:A}

A detailed derivation of Eq.~\eqref{eq:simPert} has been provided in~\cite{Mensah2020}, to which we refer the interested reader. For completeness, we report here without demonstration the explicit expression that can be used to calculate the vectors $\bm{r}_j$ at any order:
\begin{align}
\bm{r}_j \equiv \sum_{n=1}^j\mathsfbi L_{0,n}\vct p_{j-n} +
\sum_{\substack{0 < \weigh{ \mu} \le j \\ \vct \mu \neq \vct 1_j}}
\sum_{n=0}^{j-\weigh{\mu}}
\binom{|\vct \mu|}{\vct \mu }\vct s^{\vct\mu}\mtx L_{|\vct\mu|,n}\vct p_{j-n-\weigh{\mu}},
\label{eq:rk}
\end{align}
where we have introduced the multi-index notation $\vct\mu\equiv[\mu_1,\mu_2,\ldots,\mu_N]$, where $\mu_n \in \mathbb{N}$, and we have defined the multi-index 
$\vct1_j$ as having index 1 at position $j$, and 0 otherwise. The following notation for multi-index properties holds:
\begin{align}
|\vct\mu|\equiv\sum_{n=1}^N \mu_n, 
\qquad
\weigh{\mu}\equiv\sum_{n=1}^N n\mu_n,
\qquad
\binom{|\vct \mu|}{\vct \mu }\equiv\frac{|\vct\mu|!}{\displaystyle{\prod_{n=1}^N\left(\mu_n!\right)}},
\qquad
\vct s^{\vct\mu} \equiv  \prod_{n=1}^N s_n^{\mu_n}.
\end{align}
From~\eqref{eq:rk}, the explicit expressions of $\bm{r}_j$ for the first two orders are:
\begin{subequations}
\begin{align}
\varepsilon: \qquad \bm{r}_1 &\equiv  \mathsfbi{L}_{0,1}\hat{\bm{p}}_0,
\label{eq:r1}
\\
\varepsilon^2: \qquad \bm{r}_2 &\equiv (\mathsfbi{L}_{0,1} + s_1 \mathsfbi{L}_{1,0})\hat{\bm{p}}_1 + (\mathsfbi{L}_{0,2} + s_1\mathsfbi{L}_{1,1} +  s_1^2\mathsfbi{L}_{2,0})\hat{\bm{p}}_0.
\label{eq:r2}
   \end{align}%
\end{subequations}

We lastly report the explicit expression for $\mathsfbi{L}_{1,0}$, which is extensively used to define the bi-orthogonal conditions~\eqref{eq:ortoBio}. From~\eqref{eq:shorthand}, in the case in which the operator $\mathsfbi{L}$ is defined via~\eqref{eq:nep_ta}, we have
\begin{equation}
\mathsfbi{L}_{1,0} = - 2s_0 + \tau \frac{(\gamma - 1)}{\overline{\rho}} \frac{\overline{Q}}{\overline{U}}n e^{-s_0\tau}\hat{\bm{n}}_{\mathrm{ref}} \bm{\cdot} \nabla_{\mathrm{ref}},
\label{eq:L10}
\end{equation}
where $s_0$ is an eigenvalue of $L$. Note that $\mathsfbi{L}_{1,0}$ is actually a discretisation of the expression on the r.h.s. of~\eqref{eq:L10}. We again emphasise that the theory discussed in this study is valid  not only for the thermoacoustic case, for which we have reported the explicit expressions~\eqref{eq:nep_ta} and~\eqref{eq:L10}, but holds for arbitrary finite-dimensional operators $\mathsfbi{L}$.

\section{Derivation of eigenvector coefficients equations}
\label{app:B}
In this Appendix we shall prove Eq.~\eqref{eq:coeffCorr} for the specific case in which the degeneracy unfolds at first order, $d=1$, which is the relevant scenario for the applications shown in the present work. The equation is however more general and holds also for $d>1$, a proof of which is contained in~\S\ref{sec:SuppDer} of the Supplementary Material.

If the degeneracy unfolds at first order, the auxiliary eigenvalue problem  $\mathsfbi{X}_1\bm{\alpha} = s_1\bm{\alpha}$ has 2 different simple eigenvalues, $s_{1,\zeta}$ and $s_{1,\eta}$ -- see~\eqref{eq:auxPs}. Furthermore, the coefficients $\bm\alpha$ identify the basis into which the degeneracy unfolds. In this basis, the auxiliary eigenvalue problem is diagonal, so that 
\begin{equation}
    \inner{\hat{\bm{p}}_{0,\eta}^\dagger}{\bm{r}_{1,\zeta}} =    \inner{\hat{\bm{p}}_{0,\eta}^\dagger}{\mathsfbi{L}_{0,1}\hat{\bm{p}}_{0,\zeta}} = -s_{1,\zeta}\delta_{\eta,\zeta},
    \label{eq:diagional}
\end{equation}
where we used~\eqref{eq:r1} for the definition of $\bm{r}_{1,\zeta}$.

Then the eigenvector correction at first order is
\begin{align}
\begin{aligned}
    \hat{\bm{p}}_{1,\zeta} & =  -\mathsfbi{L}_{0,0}^g\left[  
    \bm{r}_{1,\zeta} + s_{1,\zeta}\mathsfbi{L}_{1,0}\hat{\bm{p}}_{0,\zeta} \right] + c_{1,\zeta,\zeta}\hat{\bm{p}}_{0,\zeta} + c_{1,\zeta,\eta}\hat{\bm{p}}_{0,\eta} = \\
    & =\hat{\bm{p}}^\bot_{1,\zeta} + c_{1,\zeta,\zeta}\hat{\bm{p}}_{0,\zeta} + c_{1,\zeta,\eta}\hat{\bm{p}}_{0,\eta},
\end{aligned}
\label{eq:p1}
\end{align}
analogous to Eq.~\eqref{eq:Fullp}.

Using the expression for $\bm{r}_2$, \eqref{eq:r2}, on branch $\zeta$, the perturbation equation at second order is
\begin{equation}
     \mathsfbi{L}_{0,0}\hat{\bm{p}}_{2,\zeta} = -(\mathsfbi{L}_{0,1} + s_{1,\zeta} \mathsfbi{L}_{1,0})\hat{\bm{p}}_{1,\zeta} - (\mathsfbi{L}_{0,2} + s_{1,\zeta}\mathsfbi{L}_{1,1} +  s_{1,\zeta}^2\mathsfbi{L}_{2,0})\hat{\bm{p}}_{0,\zeta} - s_{2,\zeta} \mathsfbi{L}_{1,0} \hat{\bm{p}}_{0,\zeta}.
\end{equation}
To be solvable, this equation needs to satisfy the conditions $\eqref{eq:2Conds3}$. We will now focus only on the second of these solvability conditions,~\eqref{eq:Fred2}, for which $\eta\neq\zeta$, and we will use it to derive an equation for the coefficients $c_{1,\zeta,\eta}$,~\eqref{eq:coeffCorr}. This condition explicitly reads
\begin{align}
    \begin{aligned}
    \inner{\hat{\bm{p}}_{0,\eta}^\dagger}{(\mathsfbi{L}_{0,1} + s_{1,\zeta} \mathsfbi{L}_{1,0})\hat{\bm{p}}_{1,\zeta} + (\mathsfbi{L}_{0,2} + s_{1,\zeta}\mathsfbi{L}_{1,1} +  s_{1,\zeta}^2\mathsfbi{L}_{2,0})\hat{\bm{p}}_{0,\zeta} + s_{2,\zeta} \mathsfbi{L}_{1,0} \hat{\bm{p}}_{0,\zeta}} = 0.
    \end{aligned}
    \label{eq:cond22}
\end{align}
We immediately notice that the last term vanishes since  $\inner{\hat{\bm{p}}_{0,\eta}^\dagger}{\mathsfbi{L}_{1,0} \hat{\bm{p}}_{0,\zeta} } = 0 $ for $\eta\neq\zeta$, from the bi-orthogonality conditions~\eqref{eq:ortoBio}. We then expand the term containing $\hat{\bm{p}}_{1,\zeta}$ using its explicit expression, from~\eqref{eq:p1}. Equation~\eqref{eq:cond22} then becomes
\begin{align}
    \begin{aligned}
    &\inner{\hat{\bm{p}}_{0,\eta}^\dagger}{(\mathsfbi{L}_{0,1} + s_{1,\zeta} \mathsfbi{L}_{1,0})\hat{\bm{p}}^\bot_{1,\zeta} + (\mathsfbi{L}_{0,2} + s_{1,\zeta}\mathsfbi{L}_{1,1} +  s_{1,\zeta}^2\mathsfbi{L}_{2,0})\hat{\bm{p}}_{0,\zeta}}~+  \\
     +~&c_{1,\zeta,\zeta}\inner{\hat{\bm{p}}_{0,\eta}^\dagger}{(\mathsfbi{L}_{0,1} + s_{1,\zeta} \mathsfbi{L}_{1,0})\hat{\bm{p}}_{0,\zeta}} + c_{1,\zeta,\eta}\inner{\hat{\bm{p}}_{0,\eta}^\dagger}{(\mathsfbi{L}_{0,1} + s_{1,\zeta} \mathsfbi{L}_{1,0})\hat{\bm{p}}_{0,\eta}}
    = 0.
    \end{aligned}
    \label{eq:cond22s1}
\end{align}
We note that the right-side argument of the inner product in the first line of this equation is formally identical to the definition of $\bm{r}_2$ from~\eqref{eq:r2}, with the difference that $\bm{p}_1$ is replaced by $\bm{p}_1^\bot$. This suggests the definition of
\begin{equation}
    \bm{r}_{2,\zeta}^\bot \equiv (\mathsfbi{L}_{0,1} + s_{1,\zeta} \mathsfbi{L}_{1,0})\hat{\bm{p}}^\bot_{1,\zeta} + (\mathsfbi{L}_{0,2} + s_{1,\zeta}\mathsfbi{L}_{1,1} +  s_{1,\zeta}^2\mathsfbi{L}_{2,0})\hat{\bm{p}}_{0,\zeta}.
    \label{eq:part1}
\end{equation}
We then tackle the inner product multiplying $c_{1,\zeta,\zeta}$ in Eq~\eqref{eq:cond22s1}. Using the linearity of the inner product, we can split it into two parts, both of which vanish:
\begin{equation}
    \inner{\hat{\bm{p}}_{0,\eta}^\dagger}{\mathsfbi{L}_{0,1}\hat{\bm{p}}_{0,\zeta}} + s_{1,\zeta}     \inner{\hat{\bm{p}}_{0,\eta}^\dagger}{\mathsfbi{L}_{1,0}\hat{\bm{p}}_{0,\zeta}} = 0.
    \label{eq:part2}
\end{equation}
The first term vanishes because of the diagonalization of the basis~\eqref{eq:diagional}, and the second term vanishes because of the bi-orthogonality condition~\eqref{eq:ortoBio}. This proves that the coefficients $c_{1,\zeta,\zeta}$ do not affect the solution of the perturbation equations at the next orders. These coefficients can be uniquely determined if a normalisation condition is imposed on the eigenvector, however, this is not discussed here.
Finally, we consider the inner product multiplying $c_{1,\zeta,\eta}$ in Eq.~\eqref{eq:cond22s1}, which simplifies to
\begin{equation}
    \inner{\hat{\bm{p}}_{0,\eta}^\dagger}{\mathsfbi{L}_{0,1}\hat{\bm{p}}_{0,\eta}} + s_{1,\zeta}     \inner{\hat{\bm{p}}_{0,\eta}^\dagger}{\mathsfbi{L}_{1,0}\hat{\bm{p}}_{0,\eta}} = -s_{1,\eta} + s_{1,\zeta},
    \label{eq:part3}
\end{equation}
where we have used again~\eqref{eq:diagional} for the first term, and the bi-orthogonality condition~\eqref{eq:ortoBio} on branch $\eta$ for the second term.

Substituting the simplified expressions~\eqref{eq:part1}, \eqref{eq:part2}, \eqref{eq:part3} into Eq.~\eqref{eq:cond22s1}, we finally have

\begin{align}
    \begin{aligned}
    \inner{\hat{\bm{p}}_{0,\eta}^\dagger}{\bm{r}_{2,\zeta}^\bot } + c_{1,\zeta,\eta}\left(-s_{1,\eta} + s_{1,\zeta}\right)
    = 0,
    \end{aligned}
    \label{eq:cond22s2}
\end{align}
from which an explicit expression for the coefficients $c_{1,\zeta,\eta}$ follows:
\begin{equation}
    c_{1,\zeta,\eta}
    = \frac{\inner{\hat{\bm{p}}_{0,\eta}^\dagger}{\bm{r}_{2,\zeta}^\bot }}{s_{1,\eta} -s_{1,\zeta}} \quad \mbox{for } \eta\neq\zeta,
\end{equation}
which is~\eqref{eq:coeffCorr} for $d=1$ and $j=1$. This proof can be extended to higher orders $n$ with appropriate definitions of the vectors $\bm{r}_{n,\zeta}^\bot$, which, analogously to~\eqref{eq:part1}, are obtained from the definition of $\bm{r}_{n,\zeta}$ by replacing $\hat{\bm{p}}_{n-1,\zeta}$ with $\hat{\bm{p}}_{n-1,\zeta}^\bot$.

\bibliographystyle{jfm}
%\bibliography{biblio}

\end{document}